\documentclass[aps,prr,twocolumn,superscriptaddress]{revtex4}
\usepackage{graphicx}
\usepackage{dcolumn}
\usepackage{bm}
\usepackage{physics}
\usepackage{amsmath}
\usepackage{amssymb}
\usepackage{array}
\usepackage{color}
\newcolumntype{P}[1]{>{\centering\arraybackslash}p{#1}}
\usepackage[colorlinks=true, breaklinks=true, linkcolor=blue, citecolor=blue, urlcolor=blue]{hyperref}

\newcommand{\geneva}{Department of Quantum Matter Physics, University of Geneva, Quai Ernest-Ansermet 24, 1211 Geneva, Switzerland}

\usepackage{ifthen}
\usepackage{booktabs}
\usepackage{multirow}

\begin{document}

\title{Exploring Frustration Effects of Strongly Interacting Bosons via the Hall Response}

\date{\today}

\begin{abstract}
We investigate the Hall response of hardcore bosonic atoms on a triangular ladder in a magnetic field. To access the dynamical properties of this many-body quantum system, we employ matrix product states numerical methods.
We show that the behavior of the Hall polarization, both in its saturation value and in the short-time dynamics, correlates with the features of the underlying phase diagram, which stem from the interplay of interactions and geometric frustration. 
This paves the way to employ the Hall response as a sensitive probe of many-body chiral quantum phases in strongly correlated materials.
\end{abstract}
\author{Catalin-Mihai Halati}
\affiliation{\geneva}
\author{Thierry Giamarchi}
\affiliation{\geneva}
\maketitle

\section{Introduction}

Strongly correlated quantum systems exhibit an intricate interplay between various energy scales and degrees of freedom. This makes the systems susceptible to perturbations and driving, opening the opportunity for probing and controlling their dynamics to investigate novel quantum behaviors \cite{BasovHsieh2017}.
In particular, in low dimensional frustrated quantum systems, the presence of different interactions which cannot be simultaneously satisfied can give rise to quantum phases with exotic ordering \cite{MoessnerRamirez2006, LacroixMila2011, Balents2010, Starykh2015}.
Due to advancements in the field of of ultracold atoms in optical lattices \cite{LewensteinSen2007,BlochZwerger2008, CiracZoller2012, GoldmanZoller2016, HofstetterQin2018, SchaeferTakahashi2020}, such frustrated triangular geometries have been the focus of recent experimental studies \cite{BeckerSengstock2010,StruckSengstock2011, YangSchauss2021, MongkolkiattichaiSchauss2023, XuGreiner2023, LebratGreiner2024, PrichardBakr2024, LiJia2023}.

One of the most important transport measurements over time has been the Hall effect. It has proven to have implications way beyond the naive derivation relating its value to the number of carriers \cite{Ziman1972}.
In the case of non-interacting systems, the Hall response has been related to topological invariants of quantum systems and the curvature of Fermi surface \cite{XiaoNiu2010,ThoulessNijs1982}.
At large magnetic fields, phenomena such as the Integer Quantum Hall effect \cite{vonKlitzing1980, vonKlitzing1986} have found their place in metrology. 
In contrast, many open questions remain regarding the deep impact of interactions on the Hall effect. 
When the magnetic field is very large, the existence of a gap between Landau levels leads to the appearance of the Fractional Quantum Hall 
effect \cite{TsuiGossard1982, StormerGossard1999} and to fractional excitations \cite{Laughlin1983}, a direct consequence 
of the strong correlations existing in the system. 
When the field is small, and a gap does not protect the excitations in the bulk of the material, the situation is even more mysterious. 
Theoretical studies have partly addressed this question 
for strongly correlated systems of bosons or fermions 
making use of special geometries \cite{LopatinGiamarchi2001, LeonGiamarchi2007, Auerbach2018, FilipponeGiamarchi2019}. 

Investigations for ladder systems, made of coupled one-dimensional structures, have allowed to access the Hall effect for 
interacting systems \cite{PrelovsekZotos1999, ZotosPrelovsek2000, GreschnerGiamarchi2019}. 
It particular, in the past few years it was shown that for interacting ladders the Hall imbalance and Hall voltage could be determined reliably from numerical calculations \cite{GreschnerGiamarchi2019, BuserGiamarchi2021}, leading to a \emph{universal} behavior related to the number of carriers for large values of the interactions, or interchain tunnelling. 
The ultracold atoms implementation of artificial gauge fields and synthetic dimensions \cite{DalibardOehberg2011, GoldmanSpielman2014}, and their experimental realizations \cite{AidelsburgerBloch2011, StruckWindpassinger2012,  MiyakeKetterle2013, AtalaBloch2014, ManciniFallani2015, TaiGreiner2017, GenkinaSpielman2019,ChalopinNascimbene2020, ZhouFallani2023}, led to a successful test of the predictions for a ladder of interacting fermionic atoms \cite{ZhouFallani2023}.
This paves the way to quantitative investigations of the Hall effect of correlated systems. 
Furthermore, analytical studies for a square ladder \cite{CitroOrignac2024} confirmed the universal behavior for Galilean invariant systems, and established a generic relation between the Hall resistance and charge stiffness, hinting that the Hall response should be a particularly sensitive probe for phase transitions in correlated systems. 

However, the previous studies have been done for quantum systems in which the kinetic energy does not lead by itself to frustration effects, which can lead to very different properties under a magnetic field.
Triangular ladders, either spin ones \cite{HikiharaFurusaki2008, SudanLauchli2009,FurukawaOnoda2010,FurukawaFurusaki2012,UedaOnoda2020}, or itinerant bosonic, or fermionic ones under flux \cite{MishraParamekanti2013, AnisimovasJuzeliunas2016, AnGadway2018, RomenLaeuchli2018, GreschnerMishra2019, CabedoCeli2020, LiLi2020, RoyHauke2022, HalatiGiamarchi2023, BarbieroCeli2023, BeradzeNersesyan2023, BeradzeNersesyan2023b, BaldelliBarbiero2024}, have been shown to exhibit phase transitions that have no direct equivalent in the square ladders \cite{HalatiGiamarchi2023}.
In addition, in two dimensions the Hall effect has proven to be affected by the frustration on triangular lattices \cite{LeonMillis2008, HaerterSriram2006,HaerterSriram2006b}.
Thus, it is interesting to explore the properties of the Hall effect in the novel situation of frustrated flux ladders. 

\begin{figure}[!hbtp]
\centering
\includegraphics[width=0.48\textwidth]{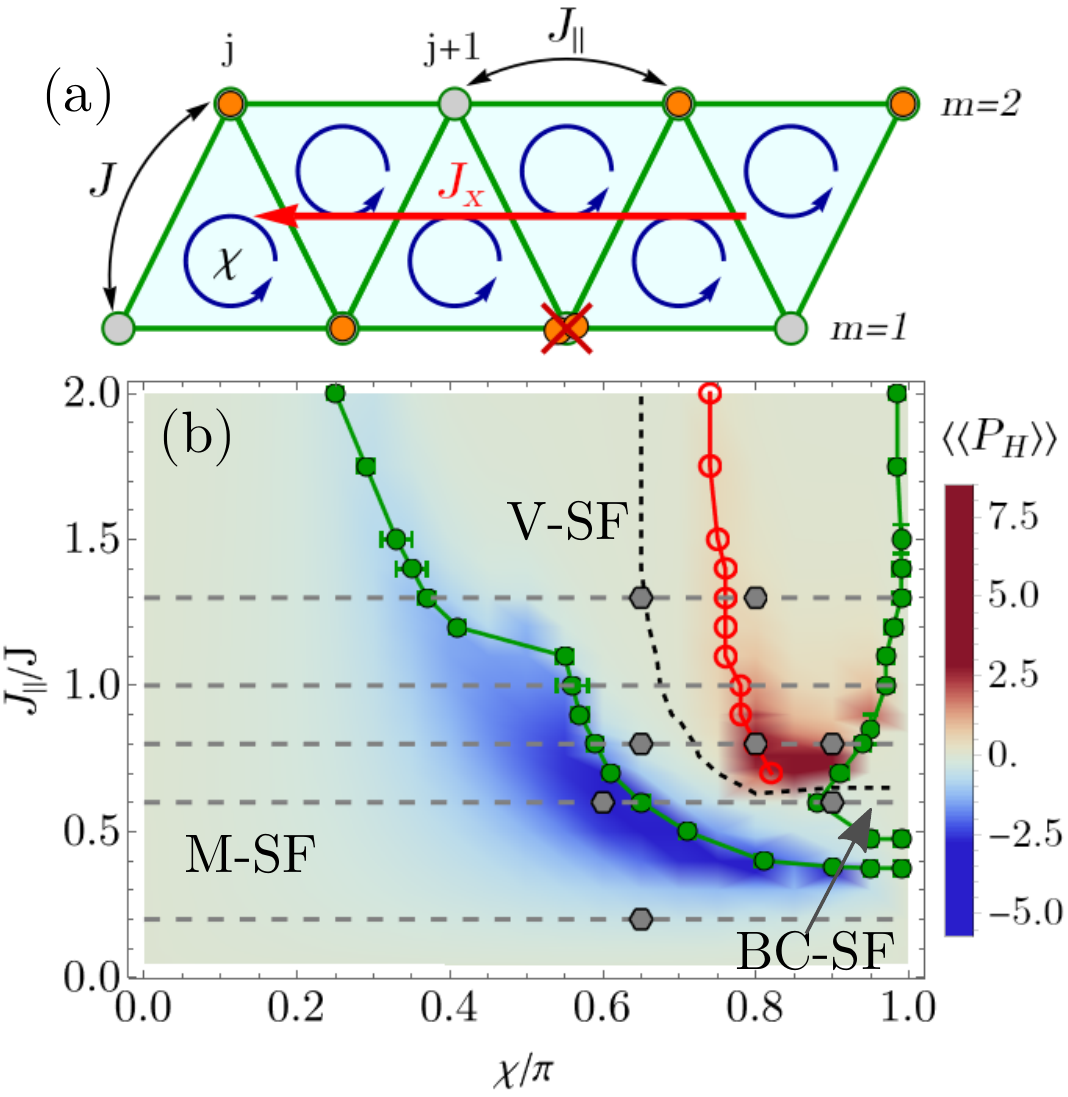}
\caption{\label{fig:sketch_PH}
(a) Sketch of the triangular ladder setup, where the legs are numbered by $m=1,2$ and the sites on each leg by $j$. The bosonic atoms can tunnel along the legs with amplitude $J_\|$ and along the rungs with amplitude $J$. We consider a hardcore on-site interaction between the atoms, and each triangular plaquette to be pierced by a flux $\chi$. We pass a current $\boldsymbol{J}_x$ through the ladder by applying a linear potential $V_x$.
(b) The value of the time-averaged Hall polarization $\langle\langle P_H\rangle\rangle$ as a function of $J_\|/J$ and $\chi$. The density of the hardcore atoms is $\rho=0.25$, the system size $L=90$ and linear potential $\mu/J=0.001$.
Green circles and curves denote the phase boundaries.
Red disks mark the maximum of the commensurate frequency peak in the rung currents.
With the dashed black curve we mark the positions where $\langle\langle P_H\rangle\rangle\approx 0$.
Gray hexagons and lines correspond to the parameters presented in Fig.~\ref{fig:PH_vs_chi} and Fig.~\ref{fig:PH_density}.}
\end{figure}

In this work, we undertake such a study to understand the behavior of the Hall response in the presence of strong interactions. We consider the case of a triangular ladder of interacting bosonic atoms under the action of a magnetic field, as introduced in Sec.~\ref{sec:model}. 
This system allows us to also harness the Hall response in order to investigate the non-trivial effects stemming from frustration. 
Employing extensive numerical simulations based on matrix product states, presented in Sec.~\ref{sec:results}, we show that the Hall response is extremely sensitive on the nature of the underlying quantum phases, offering the opportunity to  probe the features present throughout the phase diagram. 
Furthermore, the non-equilibrium dynamical behavior can be used to fingerprint the many-body chiral phases and the intriguing frustration effects, stemming from the interplay of hardcore interactions, magnetic flux and triangular geometry.

\section{Setup and protocol \label{sec:model}}

We study bosonic atoms with hardcore interactions confined to a triangular ladder in an artificial gauge field, as sketched in Fig.~\ref{fig:sketch_PH}(a). 
The Bose-Hubbard Hamiltonian of the system is given by \cite{HalatiGiamarchi2023}
\begin{align} 
\label{eq:Hamiltonian}
 H= & -J \sum_{j=1}^L \left( b^\dagger_{j,1}b_{j,2} + b^\dagger_{j+1,1}b_{j,2} +\text{H.c}. \right) \\
 & -J_\| \sum_{j=1}^{L-1} \left( e^{-i\chi}b^\dagger_{j,1}b_{j+1,1}+e^{i\chi} b^\dagger_{j,2}b_{j+1,2} + \text{H.c}. \right) \nonumber 
\end{align}
The bosonic annihilation and creation operators $b_{j,m}$ and $b^\dagger_{j,m}$ act at position $j$ and leg $m=1,2$. 
The atomic density is given by $\rho=N/(2L)$, where $N=\sum n_{j,m}$ is the total number of atoms and the ladder has $L$ sites on each leg.  
The tunneling has amplitude $J_\|$ along the two legs of the ladder and $J$ along the rungs. The complex factor in the hopping stems from the artificial magnetic field, with flux $\chi$ \cite{DalibardOehberg2011, GoldmanSpielman2014}. The atoms experience hardcore on-site interactions, such that at most one particle can be found in each site. We assume $\hbar=1$ in the following.
The ground state phase of this model is extremely rich \cite{HalatiGiamarchi2023}. In this work, the relevant phases are: the Meissner superfluid (M-SF), characterized by strong chiral currents on the legs and the absence of currents on the rungs; vortex superfluid (V-SF), for which the currents form a vortex pattern incommensurate with the underlying lattice; and the biased-chiral superfluid (BC-SF), specific to the triangular ladder, which breaks the $\mathbb{Z}_2$ symmetry of the model and is characterized by an equilibrium density imbalance.
Furthermore, going beyond the parameters considered in Ref.~\cite{HalatiGiamarchi2023} we highlight the existence also of vortex lattice superfluids (VL-SF) exhibiting commensurate vortex densities of $\rho_v=0.6$ and $\rho_v=0.8$. For the determination of the phase diagram and the identification of the nature of the phases see Appendix~\ref{appC}.

\begin{figure*}[!hbtp]
\centering
\includegraphics[width=0.97\textwidth]{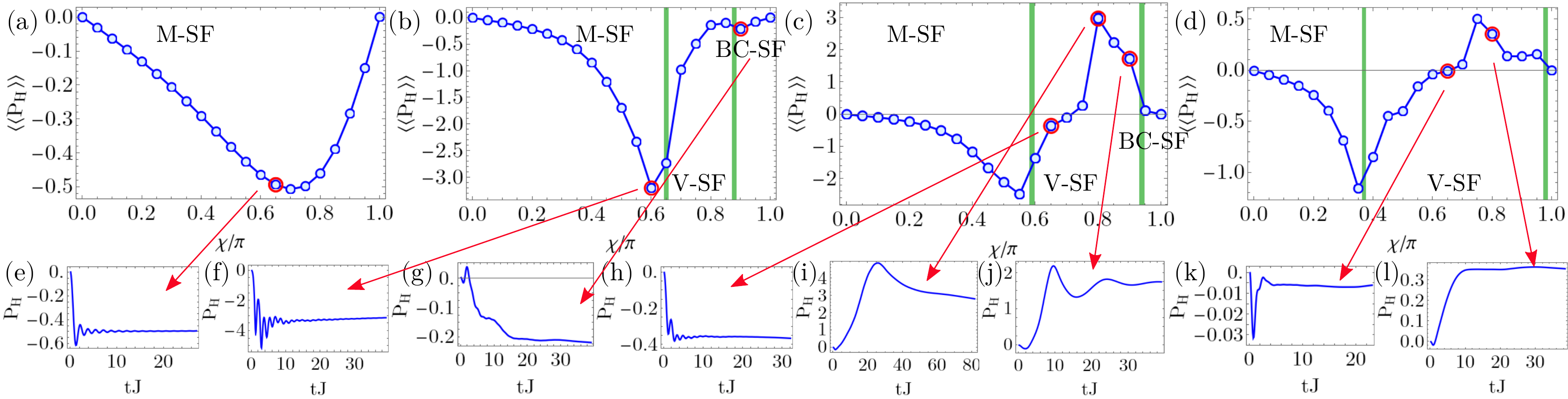}
\caption{
(a)-(d) Time-averaged Hall polarization $\langle\langle P_H\rangle\rangle$ as a function of $\chi$ for the following parameters $J_\|/J\in\{0.2,0.6,0.8,1.3\}$.
The green vertical lines mark the phase boundaries between M-SF, V-SF and BC-SF phases.
(e)-(l) Time dependence of $P_H$ for: (e) M-SF, $J_\|/J=0.2$, $\chi=0.65$; (f) M-SF, $J_\|/J=0.6$, $\chi=0.6$; (g) BC-SF, $J_\|/J=0.6$, $\chi=0.9$; (h) V-SF, $J_\|/J=0.8$, $\chi=0.65$; (i) V-SF, $J_\|/J=0.8$, $\chi=0.8$; (j) V-SF, $J_\|/J=0.8$, $\chi=0.9$; (j) V-SF, $J_\|/J=1.3$, $\chi=0.65$; (l) V-SF, $J_\|/J=1.3$, $\chi=0.8$.
The system size used is $L=90$ rungs, filling $\rho=0.25$ and the strength of the linear potential $\mu/J=0.001$.
}
\label{fig:PH_vs_chi}
\end{figure*}

We take the system out of equilibrium by quenching a linear potential in the $x$-direction, $V_x=\mu \sum_{j=1}^L\sum_{m=1}^2 \left[j+\frac{1}{2}(m-1)\right]n_{j,m}$. This protocol determines a Hall response in the system \cite{GreschnerGiamarchi2019,BuserGiamarchi2021} as in the following.
We start at time $t=0$ in the ground state of the Hamiltonian $H$, Eq.~(\ref{eq:Hamiltonian}). After the quench with the potential $V_x$, for $t>0$, a current develops in the $x$-direction, $\boldsymbol{J}_x$, and, in the presence of the magnetic flux, a density imbalance between the two legs of the ladder, $P_y=\sum_{j} \left(n_{j,1}-n_{j,2}\right)$. 
For the triangular ladder geometry the current $\boldsymbol{J}_x$ is defined as  (see Appendix~\ref{appA})
\begin{align} 
\label{eq:current_x}
\boldsymbol{J}_x=&-i\sum_{j} \Big[\frac{J}{2} \left( b^\dagger_{j,1}b_{j,2}+b^\dagger_{j,2}b_{j+1,1} -\text{H.c}.\right) \\
&+ J_\| \left(e^{-i\chi}b^\dagger_{j,1}b_{j+1,1}+e^{i\chi}b^\dagger_{j,2}b_{j+1,2}-\text{H.c.} \right)\Big]. \nonumber
\end{align}
A negative sign of $\boldsymbol{J}_x$ corresponds to a current flowing downstream with respect to the linear potential i.e. towards smaller values of the site index $j$.
The Hall response of the system is given by the ratio of the two observables, namely the Hall polarization defined as \cite{GreschnerGiamarchi2019, BuserGiamarchi2021}
\begin{align} 
\label{eq:PH}
P_H(t)=\frac{\left\langle P_y \right\rangle(t)}{\left\langle \boldsymbol{J}_x \right\rangle(t)/J}.
\end{align}
For the BC-SF phase which shows a finite value of the imbalance in equilibrium, in the numerator we consider the difference with respect to the ground state, $\left\langle P_y \right\rangle(t)-\left\langle P_y \right\rangle(0)$.
In order to compute $P_H(t)$ we simulate the time-evolution of the system starting from the ground state in the absence of the potential, using an approach based on time-dependent matrix product state methods (tMPS), as detailed in Appendix~\ref{appB}.
After the quench of the linear potential, generically, the magnitude of the density imbalance and the current $\boldsymbol{J}_x$ grows in time until the finite size effects start to influence the dynamics (as discussed in Appendix~\ref{appD}). 
However, interestingly, the Hall polarization, after a short time dynamics, exhibits a transient saturation (see Appendix~\ref{appD} and Fig.~\ref{fig:PH_vs_chi}). 
This allows us, as for previous studies on square ladders \cite{GreschnerGiamarchi2019, BuserGiamarchi2021}, to study this steady value as a function of the parameters of the model, see Fig.~\ref{fig:sketch_PH}(b). 
In the following, we denote by the Hall polarization $\langle\langle P_H\rangle\rangle$ the late time average of $P_H(t)$, where we performed the average over a time interval of at least $10/J$.

\section{\label{sec:results} Results}

We begin our analysis by investigating the dependence of $\langle\langle P_H\rangle\rangle$ as a function of the magnetic flux $\chi$ and the hopping along the legs $J_\|$, as shown in Fig.~\ref{fig:sketch_PH}(b). 
We superimpose the values computed for the Hall polarization with the phase boundaries determined in the ground state of the model, plotted with green points in Fig.~\ref{fig:sketch_PH}(b).
In this way we aim to understand the influence on the Hall response of the nature of the different chiral phases present in the ground state, which stem from the interplay between the magnetic field, interactions and frustration \cite{HalatiGiamarchi2023}.
For the results presented in Fig.~\ref{fig:sketch_PH} and Fig.~\ref{fig:PH_vs_chi} we consider a filling of $\rho=0.25$ and the following phases are present: Meissner superfluid, vortex superfluid and biased-chiral superfluid (see Appendix~\ref{appC} for the details regarding the phase diagram).

From the behavior of $\langle\langle P_H\rangle\rangle$ shown in Fig.~\ref{fig:sketch_PH}(b) we can identify several features of interest, which interestingly correlate strongly with the highlights of the ground state phase diagram.
We first observe that starting in the M-SF an increasingly stronger response is obtained as one moves towards the V-SF phase, with a large negative peak in $\langle\langle P_H\rangle\rangle$ agreeing very well with the equilibrium phase transition line between the Meissner and vortex states.
In the regime of large flux $\chi$, where the frustration effects dominate, the Hall polarization can change sign [black dashed line in Fig.~\ref{fig:sketch_PH}(b)] both in the V-SF and BC-SF phases. Intriguingly, $\langle\langle P_H\rangle\rangle$ shows also a large positive peak for $J_\|/J\gtrsim 0.7$, which matches the phase boundary between the vortex and biased phases and the presence of a commensurate dependence of the currents in the vortex phase [red symbols in Fig.~\ref{fig:sketch_PH}(b)].
In order to understand better these features of the Hall response we investigate the behavior of $\langle\langle P_H\rangle\rangle$ as a function of $\chi$ for several values of $J_\|/J$, marked by dashed gray lines in Fig.~\ref{fig:sketch_PH}(b) and shown Fig.~\ref{fig:PH_vs_chi}(a)-(d), together with the time-dependence $P_H(t)$ for several parameter sets, gray hexagons in Fig.~\ref{fig:sketch_PH}(b) and depicted in Fig.~\ref{fig:PH_vs_chi}(e)-(l).

In contrast to the square ladder, for the triangular geometry, at small values of $J_\|/J$ the Meissner superfluid phase can be found for any value of $\chi$ \cite{HalatiGiamarchi2023}. 
In the M-SF the time dependence of $P_H$ exhibits a rapid increase, followed an oscillatory regime which dampens on an time interval of around $10/J$ towards a steady value, see Fig.~\ref{fig:PH_vs_chi}(e). The dynamics is very similar to the one obtained for the Meissner phase on the square ladder \cite{GreschnerGiamarchi2019, BuserGiamarchi2021}.
If we monitor $\langle\langle P_H\rangle\rangle$ in the regime of small $J_\|/J$ we obtain a curve which vanishes at $\chi=0,\pi$ and has a maximum $\chi\approx 0.5 \pi$. However, if we increase $J_\|/J$, e.g.~in Fig.~\ref{fig:PH_vs_chi}(a) for $J_\|/J=0.2$, the height of the maximum increases and moves towards larger values of the flux. 
This behavior is determined by the proximity of the phase transition to the vortex phase, as we can see in Fig.~\ref{fig:PH_vs_chi}(b) for $J_\|/J=0.6$, where as a function of the flux we have a transition at $\chi\approx 0.65\pi$.
In the time dependence of $P_H$ for $\chi=0.6\pi$, Fig.~\ref{fig:PH_vs_chi}(f), we can see a different profile in the short-time oscillatory dynamics, compared to smaller $\chi$.
Approaching the phase transition in the M-SF $\langle\langle P_H\rangle\rangle$ exhibits a divergence-like behavior towards large negative values. 

\begin{figure}[!hbtp]
\centering
\includegraphics[width=0.48\textwidth]{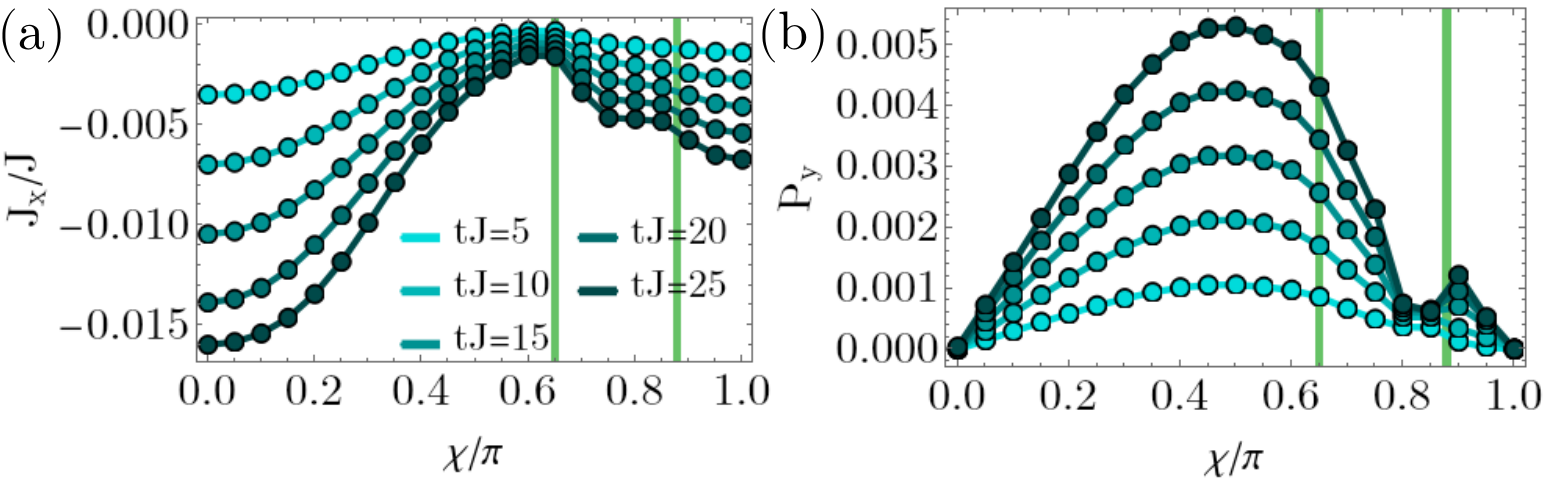}
\caption{\label{fig:obs_PT}
The dependence of the current $\boldsymbol{J}_x$ and the density imbalance $P_y$ as a function of the flux $\chi/\pi$ at different times $tJ$, for $J_\|/J=0.6$ [corresponding to the parameters in Fig.~\ref{fig:PH_vs_chi}(b)].
The green vertical lines mark the phase boundaries between M-SF, V-SF and BC-SF phases.
The values of $\boldsymbol{J}_x$ and $P_y$ are normalized with the number of rungs. 
The parameters used are $L=90$ rungs, $\rho=0.25$ and $\mu/J=0.001$.}
\end{figure}

\begin{figure}[!hbtp]
\centering
\includegraphics[width=0.38\textwidth]{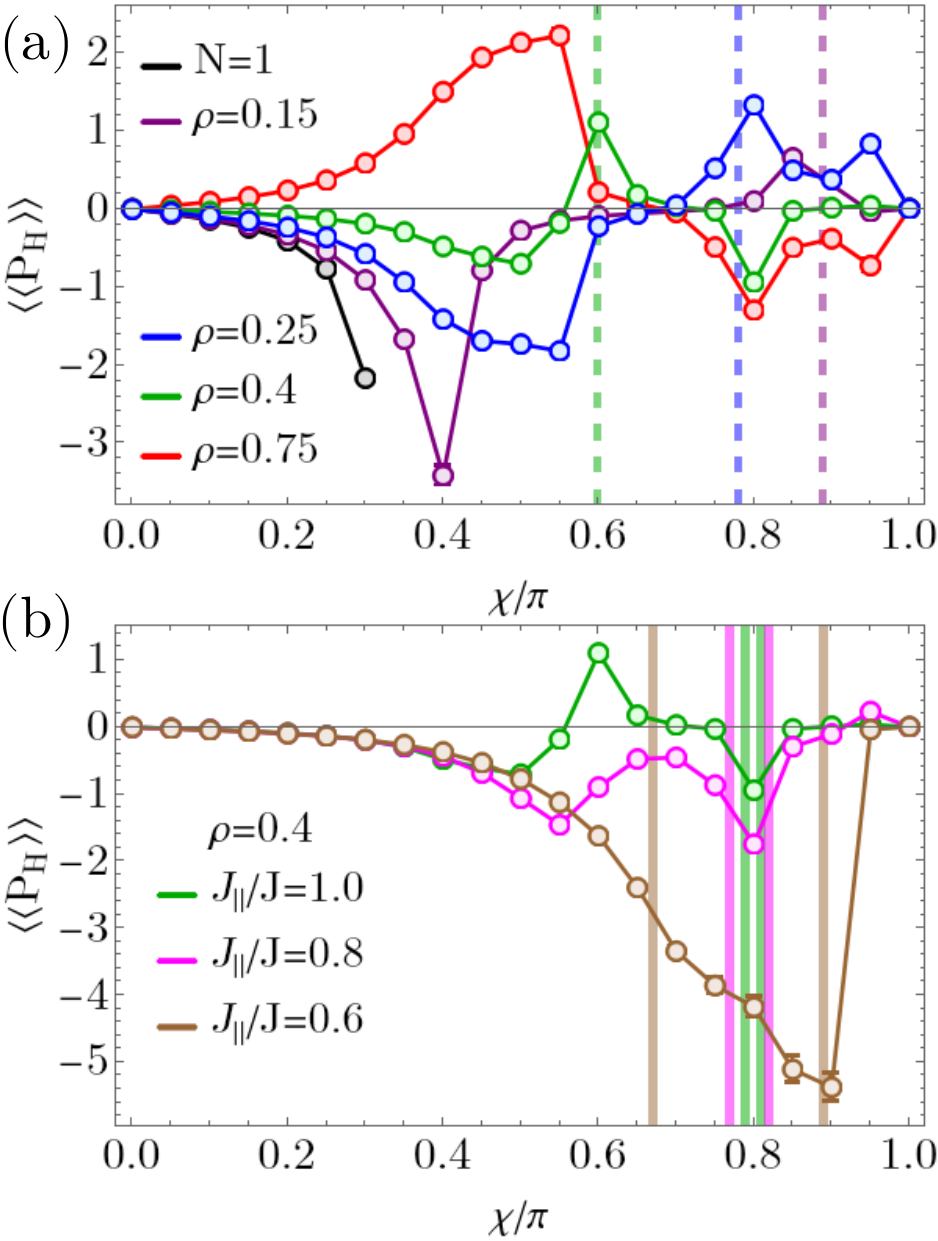}
\caption{\label{fig:PH_density}
Time-averaged Hall polarization $\langle\langle P_H\rangle\rangle$ as a function of $\chi$ for (a) different values of the filling $\rho$ at $J_\|/J=1$, (b) different values of $J_\|/J$ at $\rho=0.4$.
In (a) the vertical dashed lines mark the maximum of the commensurate peak in the rung currents, with the respective colors. In (b) the vertical lines mark the phase boundaries of the VL-SF with $\rho_v=0.8$.}
\end{figure}

The large Hall response close to the M-SF to V-SF transition is related to the presence of a minimum in the magnitude of the current $\boldsymbol{J}_x$ at the phase transition. We can see this in Fig.~\ref{fig:obs_PT}(a), where we plot $\boldsymbol{J}_x$ as a function of $\chi$ for different moments in time. We observe that the dependence of $\boldsymbol{J}_x$ on the flux exhibits an extremum at the transition threshold, which becomes more pronounced at later times.
In contrast, the minimum of $\langle\langle P_H\rangle\rangle$ in Fig.~\ref{fig:PH_vs_chi}(a) is not associated with an extrema in dependence of $\boldsymbol{J}_x$ on $\chi$, as shown in Appendix~\ref{appD}.
While in the non-interacting limit we expect the current to vanish at the transition point between M-SF and V-SF, it is surprising that even in the many-body hardcore regime we get such a strong effect in $P_H$.
This is further supported by the dependence on the atomic filling shown in Fig.~\ref{fig:PH_density}(a) for $J_\|/J=1$, with the divergent behavior being more pronounced for lower densities, in particular for single particle (black curve).

One of the highlights of the ground state of the Hamiltonian, Eq.~(\ref{eq:Hamiltonian}), is the presence of the BC-SF phase, due to frustration.
Thus, it is an interesting question to determine how the Hall polarization behaves in this phase.
As the BC-SF appears for large values of the flux, $\chi\gtrsim 0.87\pi$ in Fig.~\ref{fig:PH_vs_chi}(b) and  $\chi\gtrsim 0.93\pi$ in Fig.~\ref{fig:PH_vs_chi}(c), the magnitude of the Hall polarization is lower than in some of the other phases.
The dynamics of $P_H(t)$ in the BC-SF, Fig.~\ref{fig:PH_vs_chi}(g), shows a very distinct behavior, different than in the M-SF. At short times we observe multiple sign changes, followed by a slow evolution towards the steady value, with an intermediate plateau. In this case the main features are determined by the evolution of the imbalance $P_y$, as observed also in Fig.~\ref{fig:obs_PT}(b), with the current showing a mostly linear increase (see Appendix~\ref{appD}).
BC-SF exhibits a two-fold degenerate ground state spanned by states with density imbalances of opposite signs, however, as long as we subtract the initial imbalance, the additional imbalance $P_y$ and the current $\boldsymbol{J}_x$ are identical for both states, resulting in the same $P_H$. 

Both in the BC-SF and V-SF we obtain that the Hall polarization changes sign by varying $J_\|$ as seen in Fig.~\ref{fig:sketch_PH}(b) (sign change marked by the dashed black line).
One cause of a sign change in the Hall response is the change of the sign of the charge of the carriers, from particles to holes. 
For example this can be seen in Fig.~\ref{fig:PH_density} where $\langle\langle P_H\rangle\rangle$ has the same magnitude for $\rho=0.25$ and $\rho=0.75$ for all values of the flux, but opposite sign. This also explains why for $\rho=0.5$, in the presence of a particle-hole symmetry, the Hall polarization vanishes.
However, the situation is more complex for the change of sign at a constant density as a function of $\chi$ or $J_\|$, as we could not identify a similar argument regarding the nature of the carriers.
In the following, we analyze in more details the region of positive values of $P_H$ in the vortex phase.

In the vortex superfluid we have initially an abrupt drop in the magnitude of $\langle\langle P_H\rangle\rangle$, after entering from the Meissner phase [see Fig.~\ref{fig:PH_vs_chi}(b)-(d)], which further increases towards zero. Close to the transition the dynamics of $P_H$ looks similar to the M-SF curves [Fig.~\ref{fig:PH_vs_chi}(h) in contrast with Fig.~\ref{fig:PH_vs_chi}(e)-(f)], while for the values of the flux where $\langle\langle P_H\rangle\rangle$ is vanishing we have a rapid saturation to a small value, Fig.~\ref{fig:PH_vs_chi}(k).
After changing sign $\langle\langle P_H\rangle\rangle$ further increases to positive values, Fig.~\ref{fig:PH_vs_chi}(c)-(d).
Interestingly, the positive peak exhibits very different behavior compared to the negative one discussed previously, in the time dependence of $P_H$ we observe a significant slowing down, see Fig.~\ref{fig:PH_vs_chi}(i)-(j), with the steady plateau appearing at much longer times. 
For $J_\|/J=0.8$ the stabilization takes place after several slow oscillations [Fig.~\ref{fig:PH_vs_chi}(i)], which are absent for $J_\|/J=1.3$ [Fig.~\ref{fig:PH_vs_chi}(l)].

\begin{figure}[!hbtp]
\centering
\includegraphics[width=0.48\textwidth]{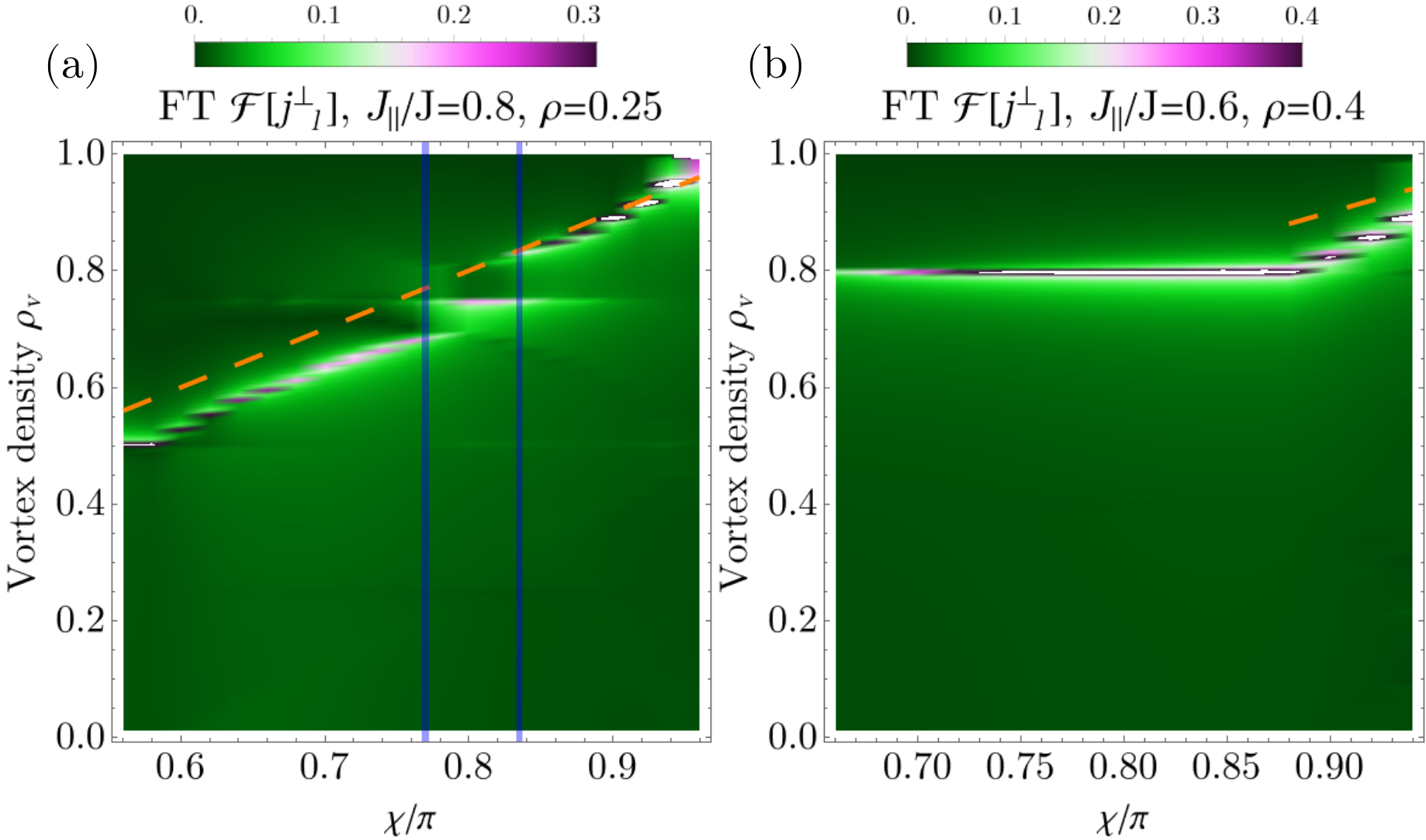}
\caption{\label{fig:vd}
The Fourier transform of the ground state local rung currents, $j^\perp_j$, as a function of the flux $\chi$ for (a) $J_\|/J = 0.8$, $\rho=0.25$, (b) $J_\|/J = 0.6$, $\rho=0.4$, and $L=120$. The vertical axis has been scaled in terms of the vortex density $\rho_v$
In panel (a) the results correspond to the vortex superfluid phase, where we expect $\rho_v=\chi/\pi$ (depicted with orange dashed lines). The vertical blue lines mark the region in which the vortex density component $\rho_v=0.75$ dominates the Fourier transform.
In panel (b) we identify a vortex lattice phase with  $\rho_v=0.8$.
}
\end{figure}

The strong positive response of the Hall polarization correlates with the proximity to the BC-SF phase, in particular for $\langle\langle P_H\rangle\rangle$ for $0.7\lesssim J_\|/J\lesssim 1.1$ [Fig.~\ref{fig:sketch_PH}(b)], and with commensurability effects occurring in the vortex phase in the limit of strong interactions.
The closeness seems indeed to be correlated to positive values of $\langle\langle P_H\rangle\rangle$ for $0.7\lesssim J_\|/J\lesssim 1.1$ [Fig.~\ref{fig:sketch_PH}(b)], however the positive peak is also present for larger values of $J_\|/J$ where the BC-SF is confined to a very narrow region close to $\chi=\pi$ [Fig.~\ref{fig:PH_vs_chi}(d) and Fig.~\ref{fig:sketch_PH}(b)].
In the following, we discuss the relation between the Hall response and the presence of a commensurate vortex density.
In Ref.~\cite{HalatiGiamarchi2023} we identified the presence of additional periodicities in the Fourier transform of the ground state rung currents, which lead to commensurate vortex densities of $\rho_v=\rho$ and $\rho_v=(1-\rho)$, beside the expected incommensurate value $\rho_v=\chi/\pi$.
Numerically, we identify the vortex density as corresponding to the frequencies for which a well-defined peak occurs in Fourier transform of the rung currents $j^\perp_l$ (see Appendix~\ref{appA} for their definition).
We show in Fig.~\ref{fig:vd}(a) the Fourier transform of the rung currents in the vortex phase for $J_\|/J=0.8$, where we observe the two peaks, one for $\rho_v=\chi/\pi$ and one for $\rho_v=(1-\rho)=0.75$ for a wide region the flux.
Upon closer inspection, we find that the Fourier response of the commensurate vortex density of $\rho_v=0.75$ can be comparable, or even dominant, with the respect to the incommensurate one [the region in between the blue vertical lines in Fig.~\ref{fig:vd}(a)]. However, interestingly, this effect does not change the nature of the V-SF phase by inducing a gap in one of the two gapless modes.
In order to establish the correlation with the behavior of the Hall polarization, we mark the position where the commensurate vortex density is dominant with red disks in Fig.~\ref{fig:sketch_PH}(b).
These follow exactly the evolution of the positive peak in $\langle\langle P_H\rangle\rangle$ in the V-SF as we increase $J_\|/J$.
Furthermore, we check this for several values of the density in Fig.~\ref{fig:PH_density}(a) and we observe a very good agreement. 
In Appendix~\ref{appD} we show that the current $\boldsymbol{J}_x$ exhibits an extremum as a function of $\chi$ resulting in the large Hall response, similar to behavior seen at the M-SF to V-SF phase transition.

We can obtain further understanding on the commensurate vortices by considering the mapping of hardcore bosons to fermions using the Jordan Wigner transformation, $b_j=\prod_{l=1}^{j-1}e^{i\pi c^\dagger_l c_l}c_j$, with $c_j$ fermionic operators. More details can be found in Appendix~\ref{sec:fermions_to_bosons}. 
For hardcore bosons on a chain the transformation results in free fermions. However, if we rewrite our Hamiltonian, Eq.~(\ref{eq:Hamiltonian}), on a chain defined along the rungs we obtain a hopping term over two sites which results in interaction term in the fermionic language, $b^\dagger_jb_{j+2}=c^\dagger_j(1-2c^\dagger_{j+1}c_{j+1})c_{j+2}$.
By varying the strength of the four-fermion term we can construct a phase diagram interpolating from free fermions on a triangular flux ladder to the hardcore bosons described by Eq.~(\ref{eq:Hamiltonian}). 
For intermediate values of the interaction, we find a vortex lattice phase with its vortex density determined by the filling, $\rho_v=1-\rho$ (see Appendix~\ref{sec:fermions_to_bosons}). This periodicity of the currents survives the phase transition to the V-SF phase towards the hardcore bosons limit.
Thus, the commensurate component of the ground state vortex density stems from the interplay of interactions and triangular geometry, determining a non-trivial behavior in the Hall polarization.

We can further emphasize the connection between a strong Hall response and the presence of commensurate current patterns in the regimes of strong magnetic fields, $\chi\gtrsim 0.5\pi$.
A strong Hall response in the presence of a commensurate current pattern seems to be characteristic only for the regime where frustration plays an important role, $\chi\gtrsim 0.5\pi$. 
We can emphasize this point by computing the Hall polarization for the genuine vortex lattice phases (VL-SF), characterized by a current pattern commensurate with the triangular ladder, which appear for an atomic filling of $\rho=0.4$. 
In this sense, we analyze the vortex lattice phases (VL-SF) appearing for an atomic filling of $\rho=0.4$ (see Appendix~\ref{appC}).
In the regime of $0.4\pi\lesssim\chi\lesssim 0.5\pi$ at $J_\|/J=1$ we have a VL-SF with $\rho_v=0.6$ , however, $\langle\langle P_H\rangle\rangle$ does not show any abrupt feature, it continues the monotonic decrease started from M-SF up to the V-SF phase [green curve in Fig.~\ref{fig:PH_density}(a)].
Interestingly, for the filling $\rho=0.4$ we observe an additional sign change in $\langle\langle P_H\rangle\rangle$ followed by a negative peak at $\chi=0.8\pi$. 
This stems from the presence of a narrow vortex lattice phase with $\rho_v=0.8$ in the ground state. 
To confirm this, in Fig.~\ref{fig:PH_density}(b) we analyze the effect of lowering $J_\|/J$ which widens the extent of the VL-SF phase. 
This significantly increases the values at which the Hall polarization stabilizes, for $J_\|/J=0.6$ this manifests itself for a wide window of flux, $0.7\pi\lesssim\chi\lesssim 0.9\pi$, corresponding to the vortex lattice phase with $\rho_v=0.8$ which we can identify in Fig.~\ref{fig:vd}. 
This further shows that the interplay of commensurate vortex current structures and frustration can have a particularly strong impact on the Hall polarization.

\section{Conclusions}

In conclusion, we have investigated the Hall response of hardcore bosons in a triangular ladder under the action of a magnetic flux.
By analyzing the features of underlying quantum states which have an important impact on the behavior of the Hall polarization, we further the understanding of the information one gathers from the Hall response in the presence of strong interactions. This analysis also establishes the saturation value of the Hall polarization, following the quench of a linear potential, as an extremely sensitive probe of the phase diagram.
We are confident that our results are relevant beyond the model studied in this work, in particular, as the Hall polarization is inversely dependent on the current $\boldsymbol{J}_x$, we expect a non-trivial behavior at phase transitions changing the nature of the equilibrium current patterns and when we have a competition between commensurate and incommensurate vortices.
In particular, our results showed large negative values of $\langle\langle P_H\rangle\rangle$ at the phase transition boundary between M-SF and V-SF. 
At large values of the flux, $P_H$ changes sign and exhibits a non-trivial behavior in connection with the commensurability effects present in the V-SF and VL-SF.
Furthermore, the non-equilibrium dynamics of the Hall polarization at short times shows behaviors particular to the initial chiral phases.
As the Hall effect has been measured experimentally with ultracold fermionic atoms for square ladders \cite{ZhouFallani2023} and recently a proposal has been put forward for the implementation of triangular flux ladders \cite{BaldelliBarbiero2024}, we expect that our results on the frustrated triangular ladder are directly relevant to current generation of ultracold atoms experiments.

\section*{ACKNOWLEDGMENTS}

We thank J.-S.~Bernier, M.~Filippone for fruitful discussions.
We acknowledge support by  the Swiss National Science Foundation under Division II grants 200020-188687 and 200020-219400. 
This research was supported in part by grant NSF PHY-1748958 to the Kavli Institute for Theoretical Physics (KITP).
The authors would like to thank the Institut Henri Poincaré (UAR 839 CNRS-Sorbonne Université) and the LabEx CARMIN (ANR-10-LABX-59-01) for their support. 

\section*{APPENDIX}

\setcounter{section}{0}
\renewcommand{\thesection}{\Alph{section}}
\renewcommand{\theequation}{A.\arabic{equation}}
\setcounter{equation}{0}

\section{\label{appA} Currents in the triangular ladder}

In this Appendix, we derive the expressions for the currents $\boldsymbol{J}_x$ and $\boldsymbol{J}_y$ flowing on the triangular ladder. We note that the expressions are equivalent to the ones obtained on the two-dimensional triangular lattice in Ref.~\cite{LeonMillis2008}.
To determine the currents we add an additional vector potential for the electric field, $\hat{A}=A_x\hat{x}+A_y\hat{y}$. This will introduce a Peierls phase in the hopping amplitudes $J_{l_1l_2}=J_{l_1l_2}\exp[-i\int_{l_1}^{l_2}\hat{A}d\hat{l}]$, where $J_{l_1l_2}$ is the hopping amplitude between two neighboring sites on the triangular ladder $l_1$ and $l_2$. Then the total currents flowing in a certain direction $\sigma$ are defined as
\begin{align} 
\label{eq:current_s}
\boldsymbol{J}_\sigma=\sum_i \left.\fdv{H}{A_\sigma(i)}\right|_{\hat{A}=0}.
\end{align}
Performing the functional derivatives we obtain
\begin{align} 
\label{eq:current_xy}
\boldsymbol{J}_x=-i&\sum_{j} \Big[\frac{J}{2} \left( b^\dagger_{j,1}b_{j,2}+b^\dagger_{j,2}b_{j+1,1} -\text{H.c}.\right) \\
&+ J_\| \left(e^{-i\chi}b^\dagger_{j,1}b_{j+1,1}+e^{i\chi}b^\dagger_{j,2}b_{j+1,2}-\text{H.c.} \right)\Big], \nonumber \\
\boldsymbol{J}_y=-i&\sum_{j} \frac{J\sqrt{3}}{2} \left( b^\dagger_{j,1}b_{j,2}+b^\dagger_{j+1,1}b_{j,2} -\text{H.c}.\right).
\end{align}

The local currents on the leg $j^\|_{j,m}$ and the rung $j^\perp_{j}$ are defined as
\begin{align}
\label{eq:localcur}
&j^\|_{j,m} = -i J_\|\left[ e^{i\chi (-1)^m} b_{j,m}^\dagger b_{j+1,m} -\text{H.c.} \right], \\
&j^\perp_{2j-1} = -i J (b_{j,1}^\dagger b_{j,2} -\text{H.c.}), \nonumber \\
&j^\perp_{2 j} = -i J (b_{j+1,1}^\dagger b_{j,2} -\text{H.c.}). \nonumber
\end{align}
Thus, we can rewrite $\boldsymbol{J}_x$ as
\begin{align} 
\label{eq:current_x}
\boldsymbol{J}_x=&\sum_{j} \left(j^\|_{j,1}+j^\|_{j,2}+\frac{1}{2}j^\perp_{2j-1} -\frac{1}{2}j^\perp_{2 j}\right).
\end{align}
We can observe that due to the triangular geometry of the ladder the current flowing in the $x$-direction has contributions also from the currents on the rungs.
We note that a positive sign $\boldsymbol{J}_x$ corresponds to a current flowing towards larger values of the site index $j$.

\setcounter{equation}{0}
\renewcommand{\theequation}{B.\arabic{equation}}

\section{\label{appB} Details on the numerical Matrix Product States methods}

The numerical results presented in this work are obtained using a
time-dependent matrix product state (tMPS) algorithm, which relies on the Trotter-Suzuki decomposition of the time evolution operator \cite{WhiteFeiguin2004, DaleyVidal2004, Schollwoeck2011}. The initial state of the evolution corresponds to the ground state of the model, Eq.~(1), and has been determined using the density matrix renormalization group (DMRG) algorithm in the matrix product state (MPS) representation \cite{White1992, Schollwoeck2011}. 
Our implementations of the algorithms make use of the ITensor Library \cite{FishmanStoudenmire2020}. 
We consider the dynamics of hardcore bosons, with a local dimension of two states, for ladders with a number of rungs of $L=90$.
For the time-dependent results presented in Sec.~\ref{sec:results} and Appendix~\ref{appD} we take a maximal bond dimension up to 400 states, which ensures that the truncation error is at most $10^{-9}$ at the final time of the evolution. 
We employ at time step of $dtJ=0.01$ and measure the observables every tenth time step.
We note that for the ground state calculations presented in Appendix~\ref{appC} we considered larger system sizes and corresponding bond dimensions, the values can be found in the captions of the figures.

\setcounter{equation}{0}
\renewcommand{\theequation}{C.\arabic{equation}}
\setcounter{figure}{0}
\renewcommand{\thefigure}{C\arabic{figure}}

\section{\label{appC} Ground state phase diagram for hardcore bosons}

\begin{figure}[!hbtp]
\centering
\includegraphics[width=0.48\textwidth]{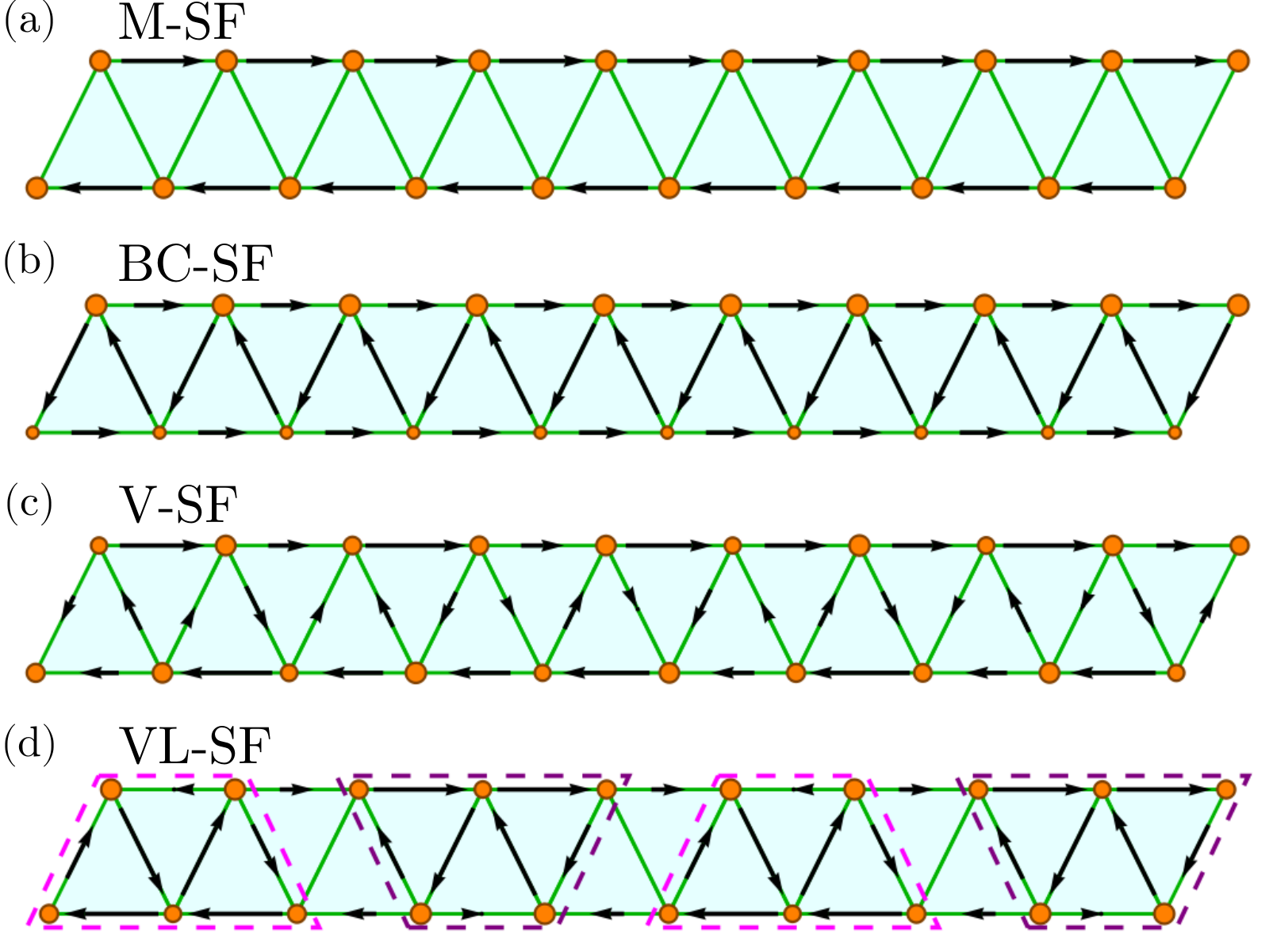}
\caption{\label{fig:currents}
The local current pattern, depicted with arrows, and
on-site densities, depicted with orange disks, obtained in the ground state for the (a) Meissner superfluid (M-SF), (b) biased chiral superfluid (BC-SF), (c) vortex superfluid phases (V-SF), and (d) vortex lattice superfluid with $\rho_v=0.8$ (VL-SF). We note that we did not normalize the local currents to the same value for the different phases presented.
In panel (d) we mark the periodic structure in the rung currents pattern.}
\end{figure}

In this Appendix, we present the results for the ground state of the model of hardcore bosons on the triangular flux ladder given in Eq.~(1).
The analysis of the numerical results used to determine the phase diagram follows Ref.~\cite{HalatiGiamarchi2023}, in which we consider similar parameters.
In order to determine the nature of the chiral phases we consider the following observables.
From the local currents, Eq.~(\ref{eq:localcur}), we compute the chiral current $J_c$ and the average rung current $J_r$ 
\begin{align}
\label{eq:cur}
&J_c = \frac{1}{2 (L-1)} \sum_j \left\langle j^\|_{j,1} - j^\|_{j,2} \right\rangle,\\ 
&J_r=\frac{1}{2 L-1} \sum_j \left|\left\langle j^\perp_{j}\right\rangle\right|. \nonumber
\end{align}
The biased phase is characterized by a finite density imbalance
\begin{align}
\label{eq:dn}
\Delta n =\frac{1}{2L}\sum_j \left(n_{j,1}-n_{j,2}\right).
\end{align}

Furthermore, we calculate the central charge $c$, which we extract from the scaling of the von Neumann entanglement entropy $S_{vN}(l)$ of subsystem of length $l$ embedded in the chain of length $L$.
For open boundary conditions the entanglement entropy for the ground state of gapless phases is given by \cite{VidalKitaev2003,CalabreseCardy2004, HolzeyWilczek1994}
\begin{equation}
\label{eq:entropy}
S_{vN}=\frac{c}{6}\log\left(\frac{L}{\pi}\sin\frac{\pi l}{L}\right)+s_1,
\end{equation}
where $s_1$ is a non-universal constant and we have neglected  corrections due to the finite size of the system \cite{AffleckLudwig1991, LaflorencieAffleck2006}.

For the parameters considered in this work we find a rich diagram \cite{HalatiGiamarchi2023} of chiral phases, whose current patterns are depicted in Fig.~\ref{fig:currents}.
We focus on the following phases in our study of the Hall response:
\begin{itemize}
    \item The \emph{Meisnner superfluid} (M-SF), characterized by strong chiral currents on the legs and vanishing currents on the rungs, see Fig.~\ref{fig:currents}(a). This state has one gapless mode and a central charge $c=1$.
    \item The \emph{biased chiral superfluid} (BC-SF), this phase breaks the $\mathbb{Z}_2$ symmetry of the ladder and is characterized by a density imbalance. The currents on the legs flow in the same direction, however their different magnitude determine a finite chiral current, while on the rungs we have a strong current in the opposite direction, Fig.~\ref{fig:currents}(b). The central charge is $c=1$.
    \item The \emph{vortex superfluid} (V-SF) has finite currents both on the legs and the rungs of the ladder. The currents determine a vortex density incommensurate with the ladder [Fig.~\ref{fig:currents}(c)], which scales linearly with the flux $\rho_v=\chi/\pi$. In the VSF both modes are gapless and $c=2$.
    \item If the vortices are pinned by the ladder and the vortex density becomes commensurate with the ladder we have a \emph{vortex lattice superfluid} (VL-SF), e.g.~Fig.~\ref{fig:currents}(d) for $\rho_v=0.8$. The central charge is $c=1$.
\end{itemize}

In order to extract the vortex density we compute the Fourier transform of the space dependence of the rung currents, $j^\perp_l$ \cite{HalatiGiamarchi2023}. In the Fourier transform we identify the frequencies corresponding to the largest peaks, which determine the vortex densities of the current pattern.

\begin{figure}[!hbtp]
\centering
\includegraphics[width=0.49\textwidth]{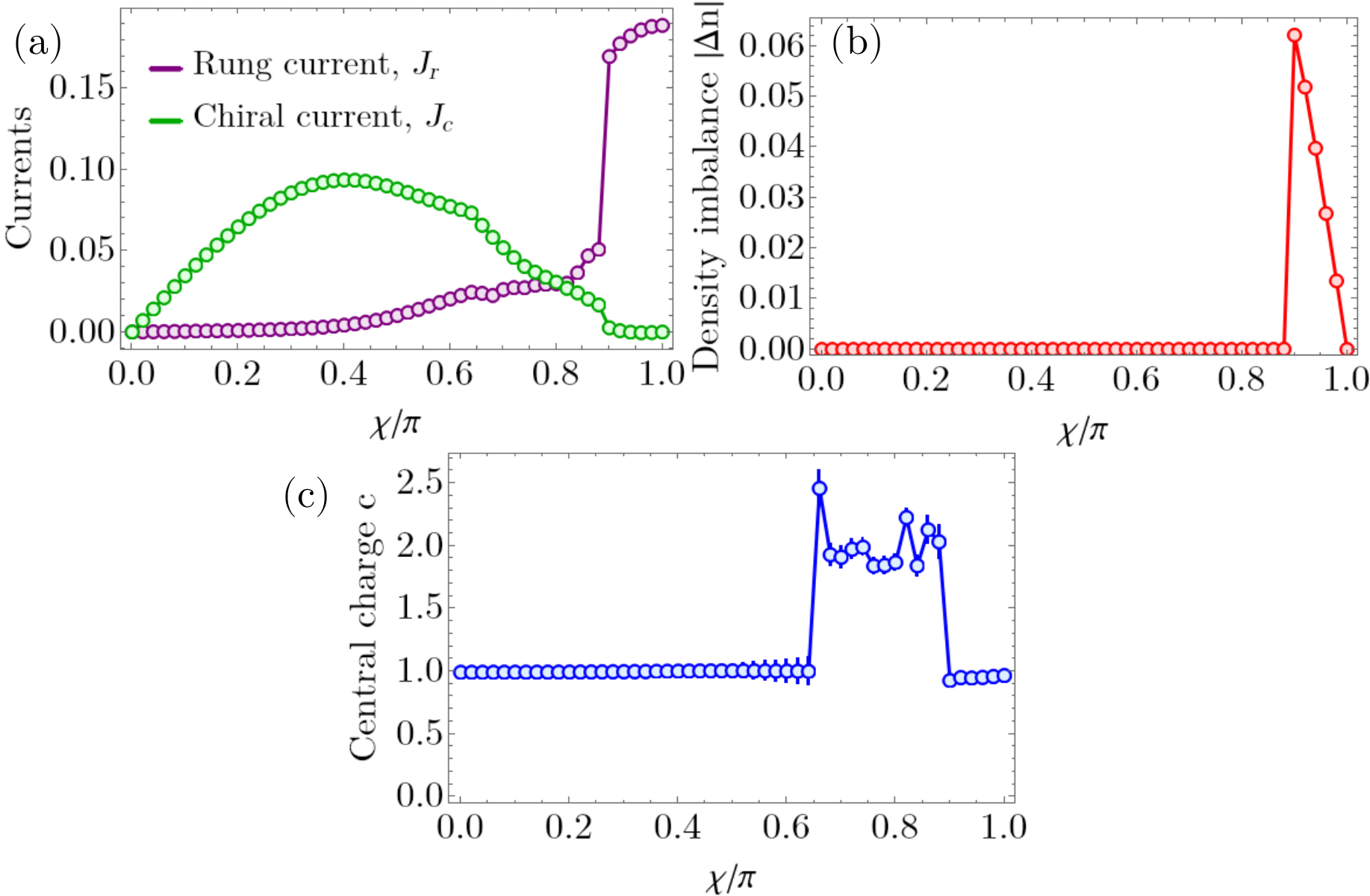}
\caption{\label{fig:gs_rho025_Jp06}
Ground state results for (a) the average rung current, $J_r$, and the chiral current, $J_c$, (b) the absolute value of the density imbalance, $|\Delta n|$, and (c) the central charge, $c$, as a function of the flux $\chi$ for $J_\|/J = 0.6$, $\rho=0.25$, $L=120$. We observe a transition from the Meissner superfluid to the vortex superfluid at $\chi\approx 0.65\pi$, and to the biased chiral superfluid at $\chi\approx 0.88\pi$. The maximal bond dimension used was $m = 750$.}
\end{figure}

\begin{figure}[!hbtp]
\centering
\includegraphics[width=0.49\textwidth]{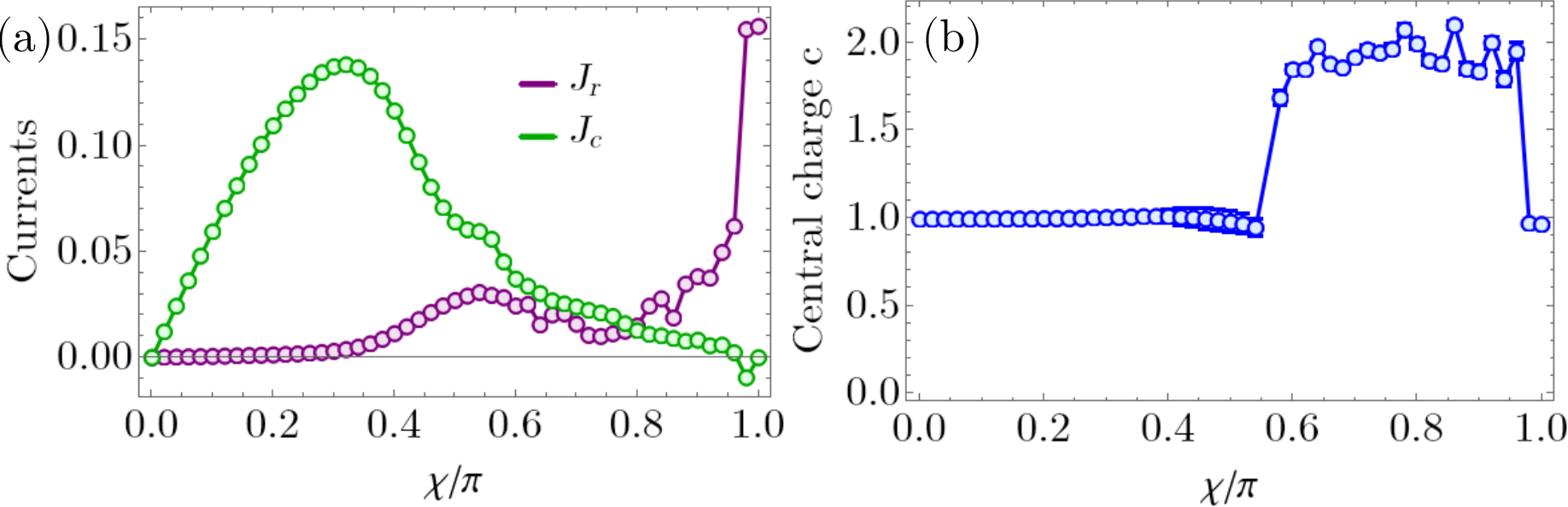}
\caption{\label{fig:gs_rho025_Jp1}
Ground state results for (a) the average rung current, $J_r$, and the chiral current, $J_c$, and (b) the central charge, $c$, as a function of the flux $\chi$ for $J_\|/J = 1$, $\rho=0.25$, $L=120$. We observe a transition from the Meissner superfluid to the vortex superfluid at $\chi\approx 0.56\pi$, and to the biased chiral superfluid at $\chi\approx 0.97\pi$. The maximal bond dimension used was $m = 750$.}
\end{figure}

For a filling of $\rho=0.25$, corresponding to the results presented in Fig.~1(b) and Fig.~2 in the main text, at small values of the flux $\chi$, or at small values of $J_\|/J$, we identify the Meissner superfluid phase. This can be concluded from the finite values of $J_c$ and vanishing $J_r$ as observed in Fig.~\ref{fig:gs_rho025_Jp06}(a) for $J_\|/J=0.6$ and Fig.~\ref{fig:gs_rho025_Jp1}(a) for $J_\|/J=1$, together with a value consistent with $c=1$ for the central charge [Fig.~\ref{fig:gs_rho025_Jp06}(c) and Fig.~\ref{fig:gs_rho025_Jp1}(b)].
By increasing the flux, or $J_\|/J$ [see the phase diagram shown in Fig.~1(b) of the main text] we enter the vortex superfluid, with finite values for both $J_r$ and $J_c$ [Fig.~\ref{fig:gs_rho025_Jp06}(a) for $J_\|/J=0.6$, Fig.~\ref{fig:gs_rho025_Jp1}(a) for $J_\|/J=1$] and $c\approx 2$ [Fig.~\ref{fig:gs_rho025_Jp06}(c) and Fig.~\ref{fig:gs_rho025_Jp1}(b)].
In the strongly frustrated regime, for $\chi$ close to $\pi$, we can find the biased chiral superfluid.
This phase has the largest extend for $J_\|/J=0.6$ and can be identified due to its finite density imbalance between the two legs of the ladder, see Fig.~\ref{fig:gs_rho025_Jp06}(b).

\begin{figure}[!hbtp]
\centering
\includegraphics[width=0.48\textwidth]{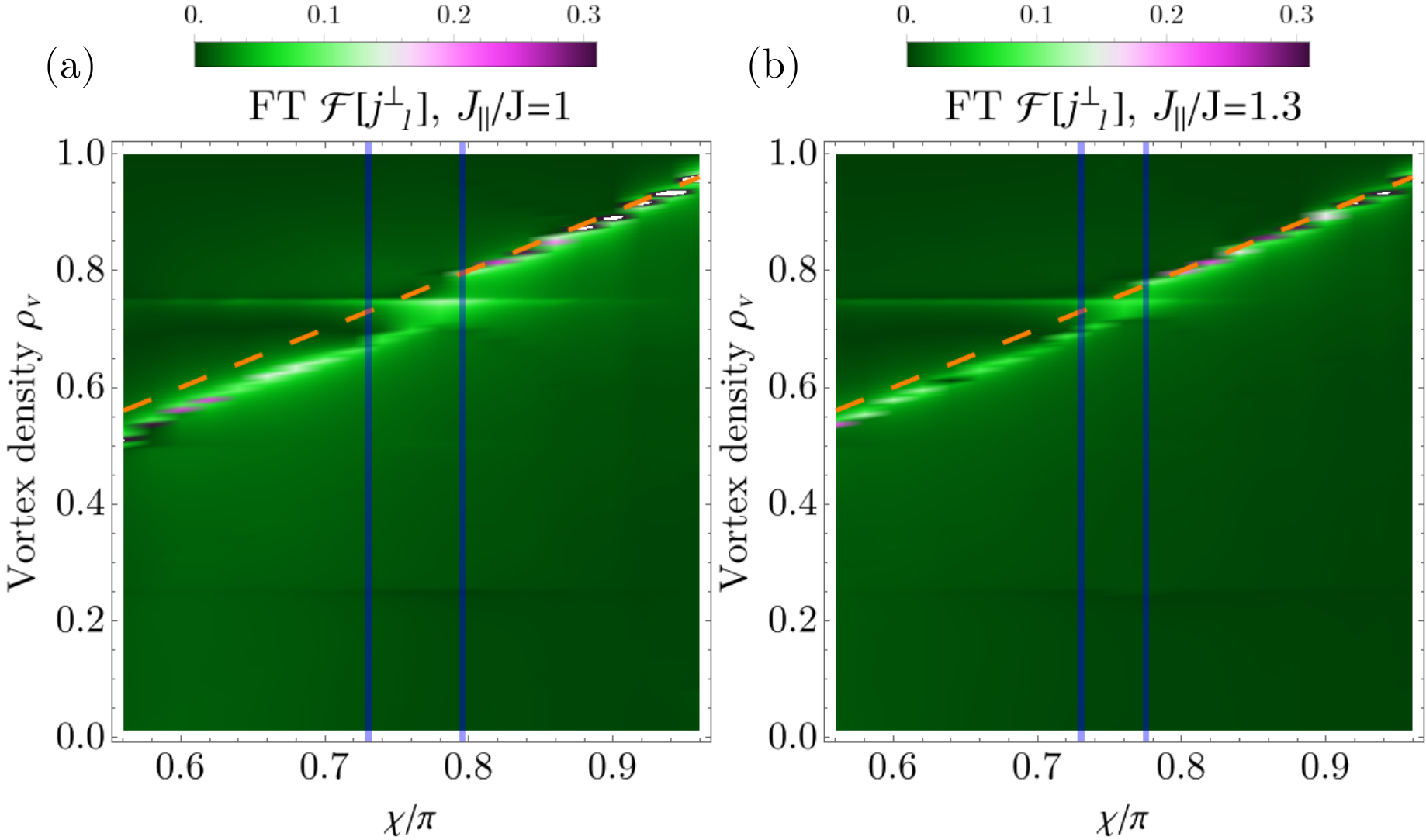}
\caption{\label{fig:vd_rho025}
The Fourier transform of the ground state local rung currents, $j^\perp_j$, as a function of the flux $\chi$ for (a) $J_\|/J = 1$, (b) $J_\|/J = 1.3$, $\rho=0.25$, $L=120$. The vertical axis has been scaled in terms of the vortex density $\rho_v$
The results correspond to the vortex superfluid phase, where we expect $\rho_v=\chi/\pi$ (depicted with orange dashed lines). The vertical blue lines mark the region in which the vortex density component $\rho_v=0.75$ dominates the Fourier transform.}
\end{figure}

\begin{figure}[!hbtp]
\centering
\includegraphics[width=0.49\textwidth]{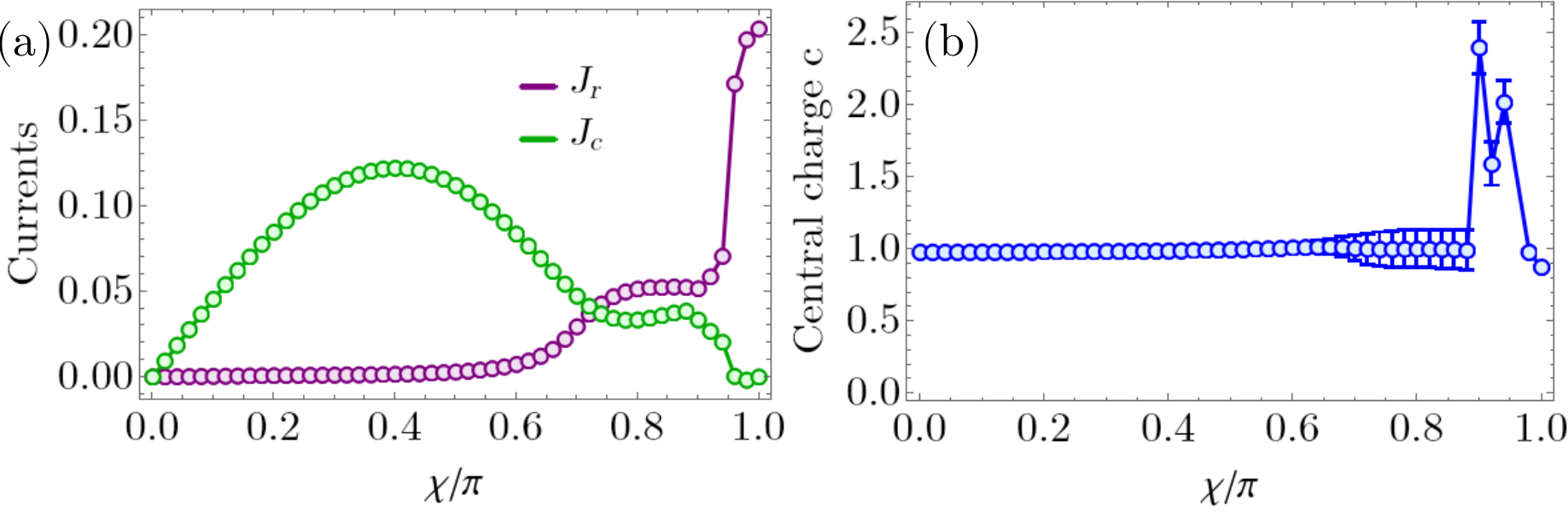}
\caption{\label{fig:gs_rho04_Jp06}
Ground state results for (a) the average rung current, $J_r$, and the chiral current, $J_c$, and (b) the central charge, $c$, as a function of the flux $\chi$ for $J_\|/J = 0.6$, $\rho=0.4$, $L=120$. We observe a transition from the Meissner superfluid to the vortex lattice superfluid with $\rho_v=0.8$ at $\chi\approx 0.67$, followed by a transition to the vortex superfluid at $\chi\approx 0.89$, and to the biased chiral superfluid at $\chi\approx 0.97$. The maximal bond dimension used was $m = 750$.}
\end{figure}

One interesting feature of the vortex superfluid phase for hardcore bosons on the triangular ladder, which we identified in Ref.~\cite{HalatiGiamarchi2023}, is the presence of a second peak in the Fourier transform of the local rung currents, besides the one corresponding to a vortex density $\rho_v\approx\chi/\pi$.
We connected the value of vortex density corresponding to the second peak to the atomic filling, $\rho_v=(1-\rho)$, as we can observe in Fig.~\ref{fig:vd_rho025}.
Surprisingly, there exists a region in which the vortex density related to the atomic filling has the stronger contribution to the Fourier transform of the rung currents [marked by vertical blue lines in Fig.~\ref{fig:vd_rho025}(a) for $J_\|/J=1$ and Fig.~\ref{fig:vd_rho025}(b) for $J_\|/J=1.3$].
However, this does not induce a pinning of the vortices and a gapped mode, as the central charge is $c\approx 2$, this can be seen in Fig.~\ref{fig:gs_rho025_Jp1}(b) for $0.73\pi\lesssim\chi\lesssim 0.79\pi$. 
Thus, the nature remains that of an incommensurate vortex phase.

\begin{figure}[!hbtp]
\centering
\includegraphics[width=0.49\textwidth]{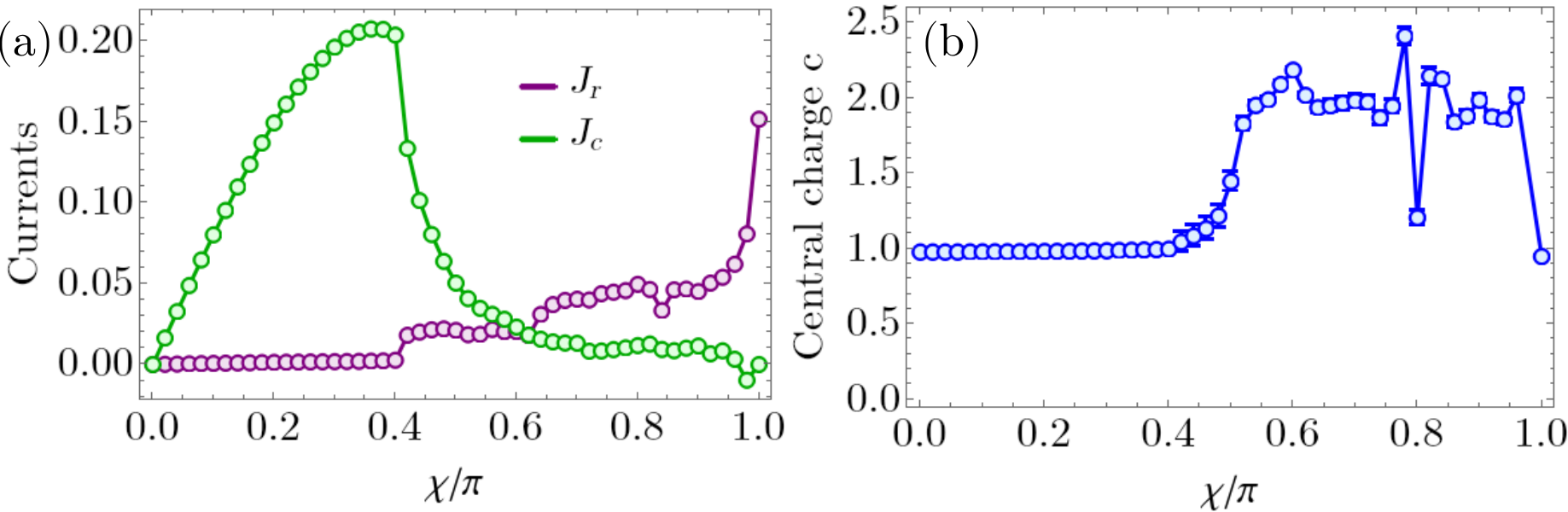}
\caption{\label{fig:gs_rho04_Jp1}
Ground state results for (a) the average rung current, $J_r$, and the chiral current, $J_c$, and (b) the central charge, $c$, as a function of the flux $\chi$ for $J_\|/J = 1$, $\rho=0.4$, $L=120$. We observe a transition from the Meissner superfluid to the vortex lattice superfluid with $\rho_v=0.6$ at $\chi\approx 0.41\pi$, followed by a transition to the vortex superfluid at $\chi\approx 0.5\pi$, between $\chi\approx 0.79$ and $\chi\approx 0.81\pi$ we have a vortex lattice superfluid with $\rho_v=0.8$, and from $\chi\approx 0.97\pi$ the biased chiral superfluid. The maximal bond dimension used was $m = 750$.}
\end{figure}

\begin{figure}[!hbtp]
\centering
\includegraphics[width=0.48\textwidth]{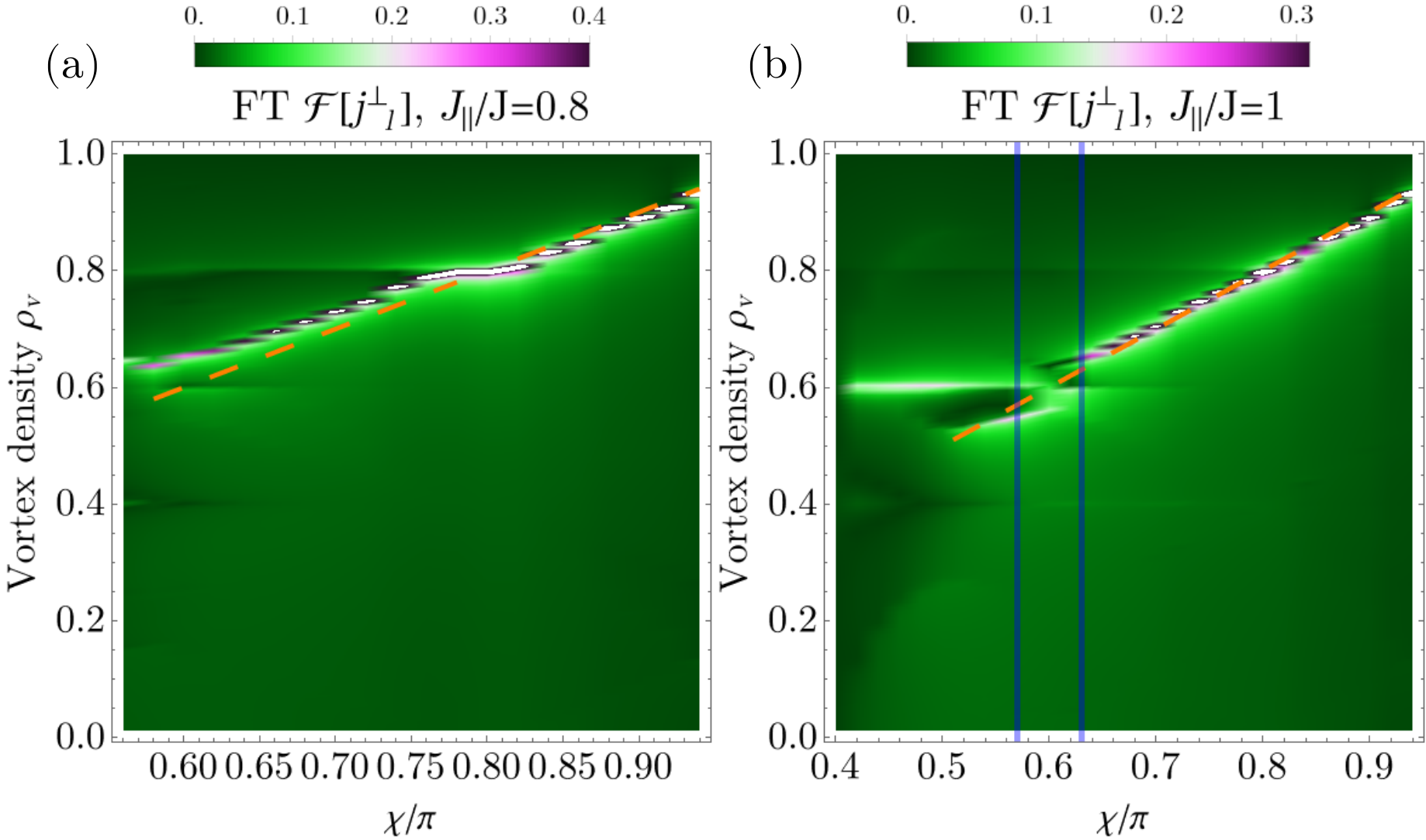}
\caption{\label{fig:vd_rho04}
The Fourier transform of the ground state local rung currents, $j^\perp_j$, as a function of the flux $\chi$ for (a) $J_\|/J = 0.8$, (b) $J_\|/J = 1$, $\rho=0.4$, $L=120$. The vertical axis has been scaled in terms of the vortex density $\rho_v$.
The orange dashed lines corresponds to the expectation of the vortex superfluid phase of $\rho_v=\chi/\pi$. In (b) the vertical blue lines mark the region in which the vortex density component $\rho_v=0.6$ dominates the Fourier transform in the VSF.}
\end{figure}

In the case of atomic filling of $\rho=0.4$, beside the three phases previously discussed, we can also find vortex phases with a commensurate vortex density, i.e.~vortex lattice superfluid, which we did not observe for the parameters of Ref.~\cite{HalatiGiamarchi2023}.
For $\rho=0.4$ and $J_\|/J=0.6$ as we increase the flux from the Meissner phase around $\chi\approx 0.67\pi$ we enter a phase with finite rung and chiral currents and $c=1$, as shown in Fig.~\ref{fig:gs_rho04_Jp06}.
If we compute the vortex density, Fig.~\ref{fig:vd}(b), we obtain for $0.67\pi\lesssim\chi\lesssim 0.89\pi$ a commensurate value of $\rho_v=0.8$. 
This implies the existence of a vortex lattice superfluid phase with $\rho_v=4/5$, the commensurate vortex pattern is shown in Fig.~\ref{fig:currents}(d).
The $\rho_v=0.8$ vortex lattice is also present for $J_\|/J=0.8$ as seen in Fig.~\ref{fig:vd_rho04}(a), however for a narrower region of the flux, $0.78\pi\lesssim\chi\lesssim 0.82\pi$.
Similarly, for $J_\|/J=1$, Fig.~\ref{fig:gs_rho04_Jp1}, we find at $\chi\approx 0.41\pi$ a transition from the Meissner phase to a vortex lattice superfluid phase with $\rho_v=0.6$.
Interestingly, if we look at the Fourier transform of the rung currents in Fig.~\ref{fig:vd_rho04}(b), we observe that between $0.41\pi\lesssim \chi\lesssim 0.5\pi$ we have only $\rho_v=0.6$, while after entering the vortex superfluid we have both $\rho_v=0.6$ and $\rho_v\approx\chi/\pi$.
And while for $0.57\pi\lesssim \chi\lesssim 0.63\pi$ $\rho_v=0.6$ has a larger weight in the Fourier transform, the state remains gapless with $c\approx 2$ [see Fig.~\ref{fig:gs_rho04_Jp1}(b)].
At larger values of the flux $\chi\gtrsim 0.7\pi$ we only observe the incommensurate vortex density. However, in the central charge around $\chi\approx 0.8\pi$ we see a deep towards $c=1$ [Fig.~\ref{fig:gs_rho04_Jp1}(b)] indicating the possibility of a narrow vortex lattice with $\rho_v=0.8$, connected to the one present at lower values of $J_\|/J$.

\setcounter{equation}{0}
\renewcommand{\theequation}{D.\arabic{equation}}
\setcounter{figure}{0}
\renewcommand{\thefigure}{D\arabic{figure}}

\section{Time dependence of the Hall Polarization\label{appD}}

In this Appendix we present additional numerical data for the behavior of the Hall polarization, $P_H$, supporting the results we discussed in Sec.~\ref{sec:results}.
In Appendix~\ref{sec:ph_msf} to Appendix~\ref{sec:ph_vlsf} we show the time evolution of $P_H$ in the different phases, from which we extracted the time-averaged $\langle\langle P_H\rangle\rangle$ depicted in the figures of Sec.~\ref{sec:results}. Furthermore, we plot the time dependence of the current $\boldsymbol{J}_x$ and imbalance $P_y$ used in the calculation of $P_H$. 
In Appendix~\ref{sec:ph_linearpot} we analyze the influence of the strength of the linear potential quenched in the protocol, which gives us a handle also on the finite size effects present in the simulations.

We note that in order to limit the influence of the finite size effects we compute the current $\boldsymbol{J}_x$ and imbalance $P_y$ for a subsystem of length $L/3$ in the middle of our system. We found that this represents a good trade-off of having enough lattice sites to compute reliably the observables and access to the dynamics of to long enough times before we observe the finite size effects of the particles being reflected by the boundaries.

\subsubsection{Hall Polarization in the Meissner superfluid phase\label{sec:ph_msf}}

\begin{figure}[!hbtp]
\centering
\includegraphics[width=0.48\textwidth]{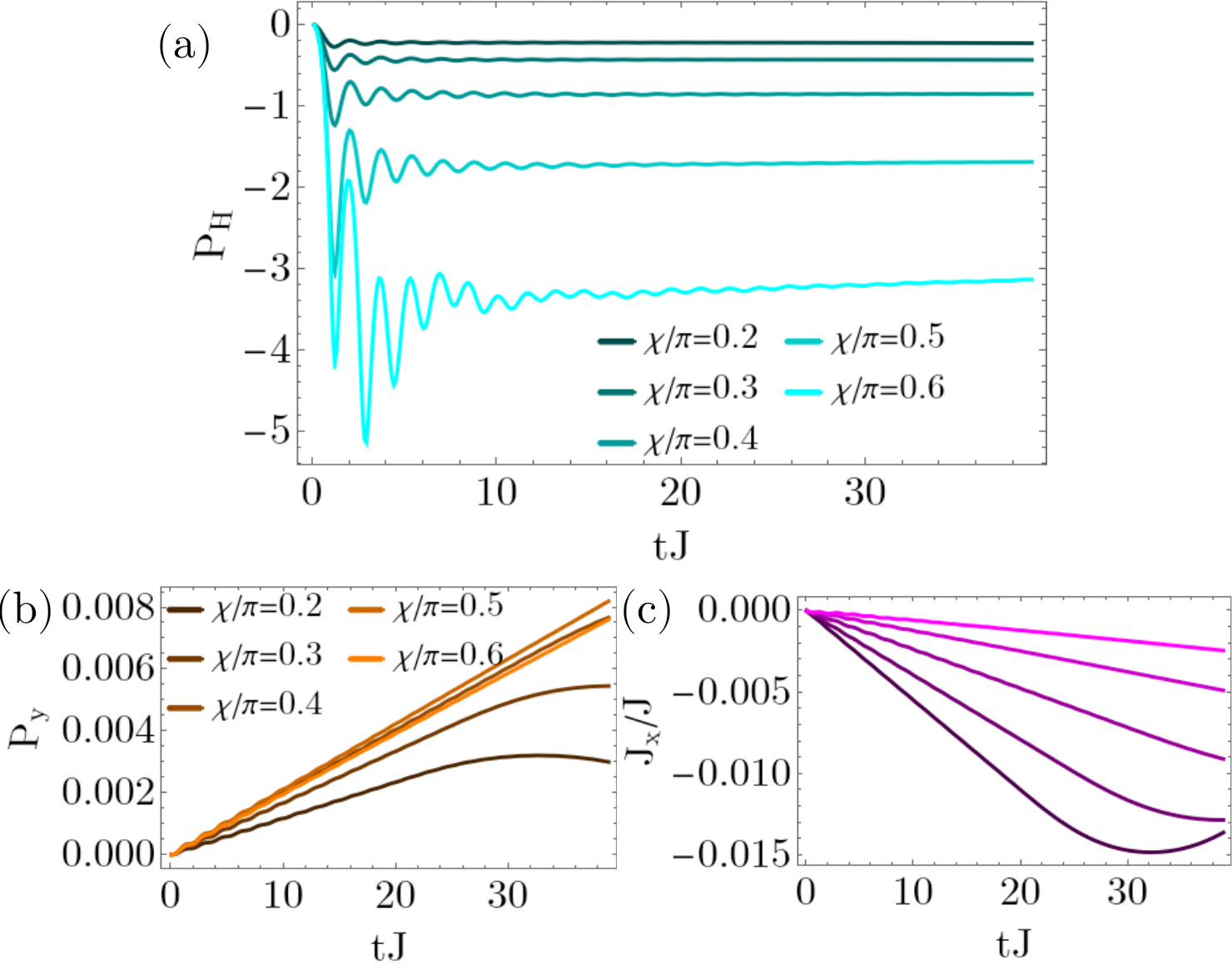}
\caption{\label{fig:pht_msf}
(a) The time evolution of $P_H$ in the Meissner superfluid phase for several values of the flux $\chi\in\{0.2\pi,0.3\pi,0.4\pi,0.5\pi,0.6\pi\}$.
The time evolution of (b) $P_y$ and (c) $\boldsymbol{J}_x$ for the same values of the flux, where the lightest shade of (b) orange and (c) magenta corresponds to $\chi=0.6\pi$ and the colors are gradually darker for the smaller values of $\chi$.
The values of $\boldsymbol{J}_x$ and $P_y$ are normalized with the number of rungs. 
The parameters used are $J_\|/J=0.6$, $L=90$ rungs, $\rho=0.25$ and $\mu/J=0.001$.
}
\end{figure}

\begin{figure}[!hbtp]
\centering
\includegraphics[width=0.48\textwidth]{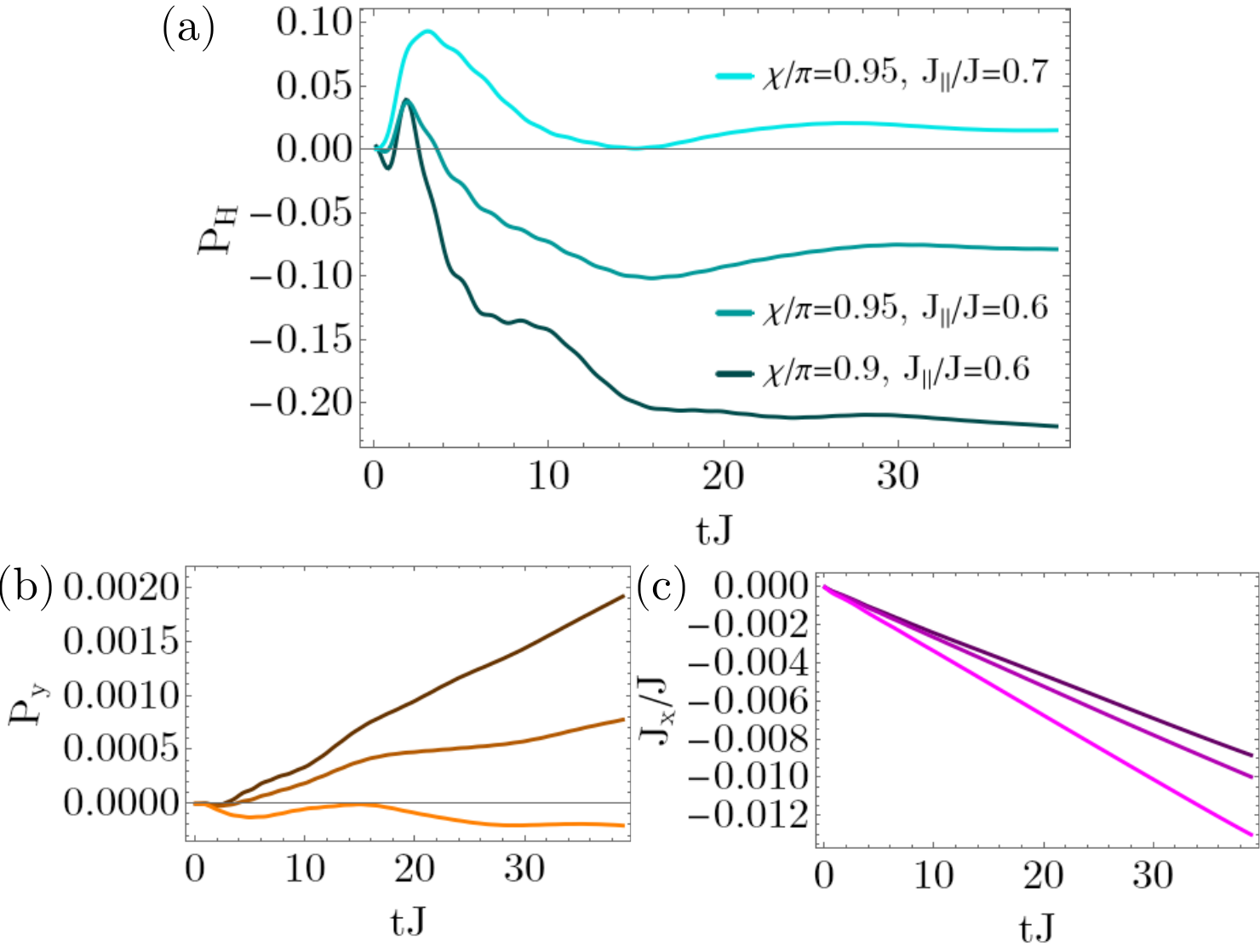}
\caption{\label{fig:pht_bcsf}
(a) The time evolution of $P_H$ in the biased chiral superfluid phase for several values of the flux $\chi=0.95\pi$, $J_\|/J=0.7$, and $\chi\in\{0.9\pi,0.95\pi\}$, $J_\|/J=0.6$.
The time evolution of (b) $P_y$ and (c) $\boldsymbol{J}_x$ for the same values of the flux and $J_\|$, where the lightest shade of (b) orange and (c) magenta corresponds to the lightest shade of cyan in (a).
The values of $\boldsymbol{J}_x$ and $P_y$ are normalized with the number of rungs.
The parameters used are $L=90$ rungs, $\rho=0.25$ and $\mu/J=0.001$.
}
\end{figure}

In Fig.~\ref{fig:pht_msf}(a) we plot the dynamics of the Hall polarization in the Meissner phase. We observe for all values of the flux a well-defined plateau occurring after the initial increase and damped oscillations.
As we increase $\chi$ towards the transition threshold to the vortex phase at $\chi\approx0.65\pi$, both the value of the late-time plateau increases (as discussed Sec.~\ref{sec:results}) and the amplitude of the oscillations.
The large values of $P_H$ close to the phase transition stems from the decrease of the magnitude of the current $\boldsymbol{J}_x$. We can see in Fig.~\ref{fig:pht_msf}(c) that the smallest values of the current correspond to $\chi=0.6\pi$ (lightest shade of magenta), while the imbalance $P_y$ has similar values for $\chi=0.4\pi$, $\chi=0.5\pi$ and $\chi=0.6\pi$ [Fig.~\ref{fig:pht_msf}(b)].

For $\chi=0.2\pi$ we see in both $P_y$ and $\boldsymbol{J}_x$ [Fig.~\ref{fig:pht_msf}(b)-(c)] that around $tJ\approx 30$ a change in the monotony occurs, we attribute this to finite size effects. However, $P_H$ shows a robust plateau up to these times, such that we can use this behavior to determine the longest times that we can use in the time averaged quantity $\langle\langle P_H\rangle\rangle$. We can also observe that for the larger values of the flux the extremum of the currents occurs at even longer times.

\subsubsection{Hall Polarization in the biased chiral superfluid phase\label{sec:ph_bcsf}}

The time evolution of $P_H$ in the biased chiral superfluid, shown in Fig.~\ref{fig:pht_bcsf}(a), exhibits a very different dynamics compared to the Meissner phase.
The saturation process is much slower, only after $tJ\approx15$ we observe a constant-like value, but which still has low frequency and small amplitude oscillations on top [in contrast to the oscillations of the M-SF seen in Fig.~\ref{fig:pht_msf}(a)].
This behavior seems to stem from the dynamics of the density imbalance $P_y$, as seen in Fig.~\ref{fig:pht_bcsf}(b), which shows oscillations with a small frequency, while the current, Fig.~\ref{fig:pht_bcsf}(c), has a mostly linear dependence on time.
Furthermore, in the BC-SF we obtain both negative and positive values of $P_H$, a sign change happening for $\chi=0.95\pi$ as we increase $J_\|/J=0.6$ to $J_\|/J=0.7$ [see Fig.~\ref{fig:pht_bcsf}(a)].

\subsubsection{Hall Polarization in the vortex superfluid phase\label{sec:ph_vsf}}

\begin{figure}[!hbtp]
\centering
\includegraphics[width=0.49\textwidth]{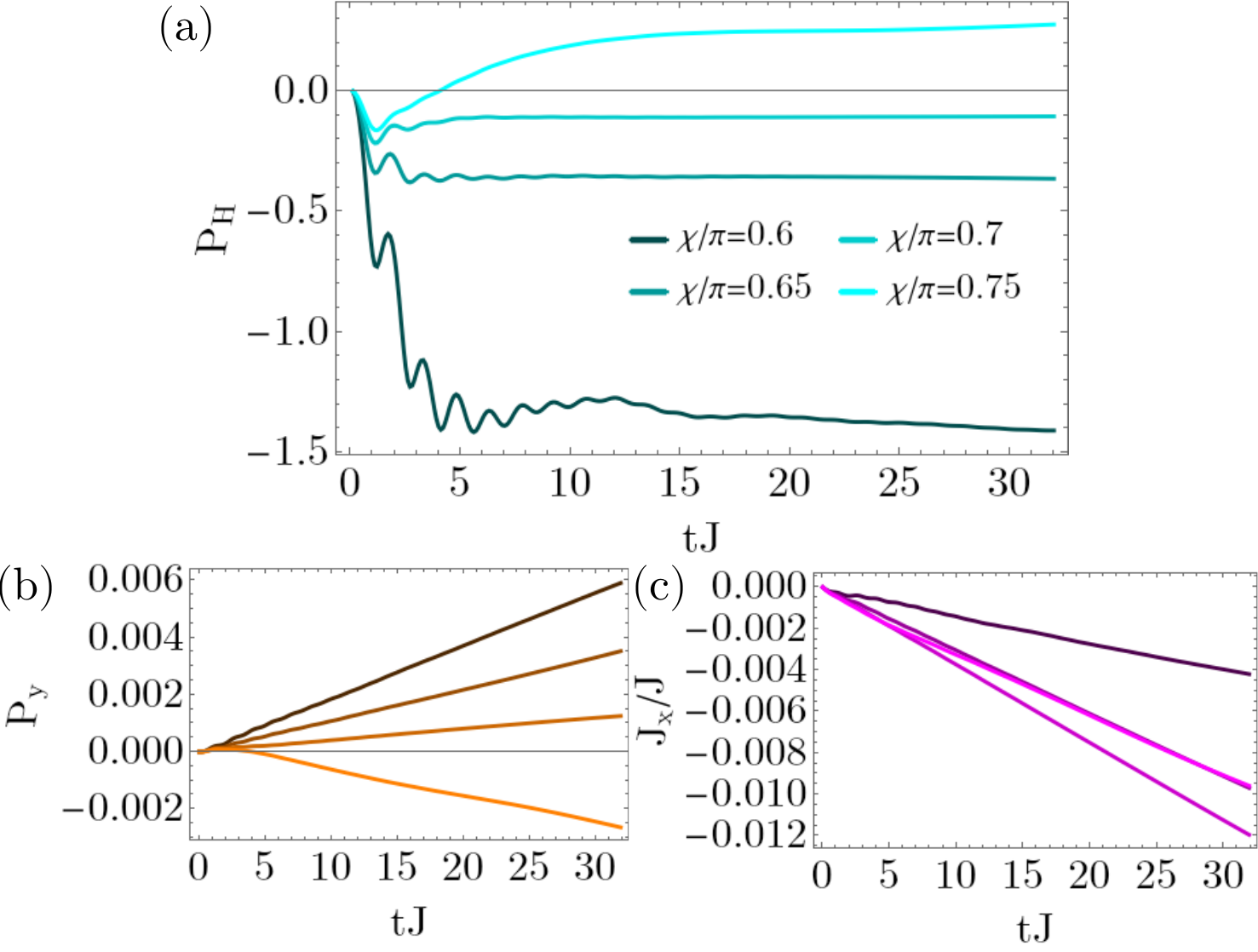}
\caption{\label{fig:pht_vsf1}
(a) The time evolution of $P_H$ in the vortex superfluid phase for several values of the flux $\chi\in\{0.6\pi,0.65\pi,0.7\pi,0.75\pi\}$.
The time evolution of (b) $P_y$ and (c) $\boldsymbol{J}_x$ for the same values of the flux and $J_\|$, where the lightest shade of (b) orange and (c) magenta corresponds to the lightest shade of cyan in (a).
The values of $\boldsymbol{J}_x$ and $P_y$ are normalized with the number of rungs.
The parameters used are $J_\|/J=0.8$, $L=90$ rungs, $\rho=0.25$ and $\mu/J=0.001$.
}
\end{figure}

\begin{figure}[!hbtp]
\centering
\includegraphics[width=0.49\textwidth]{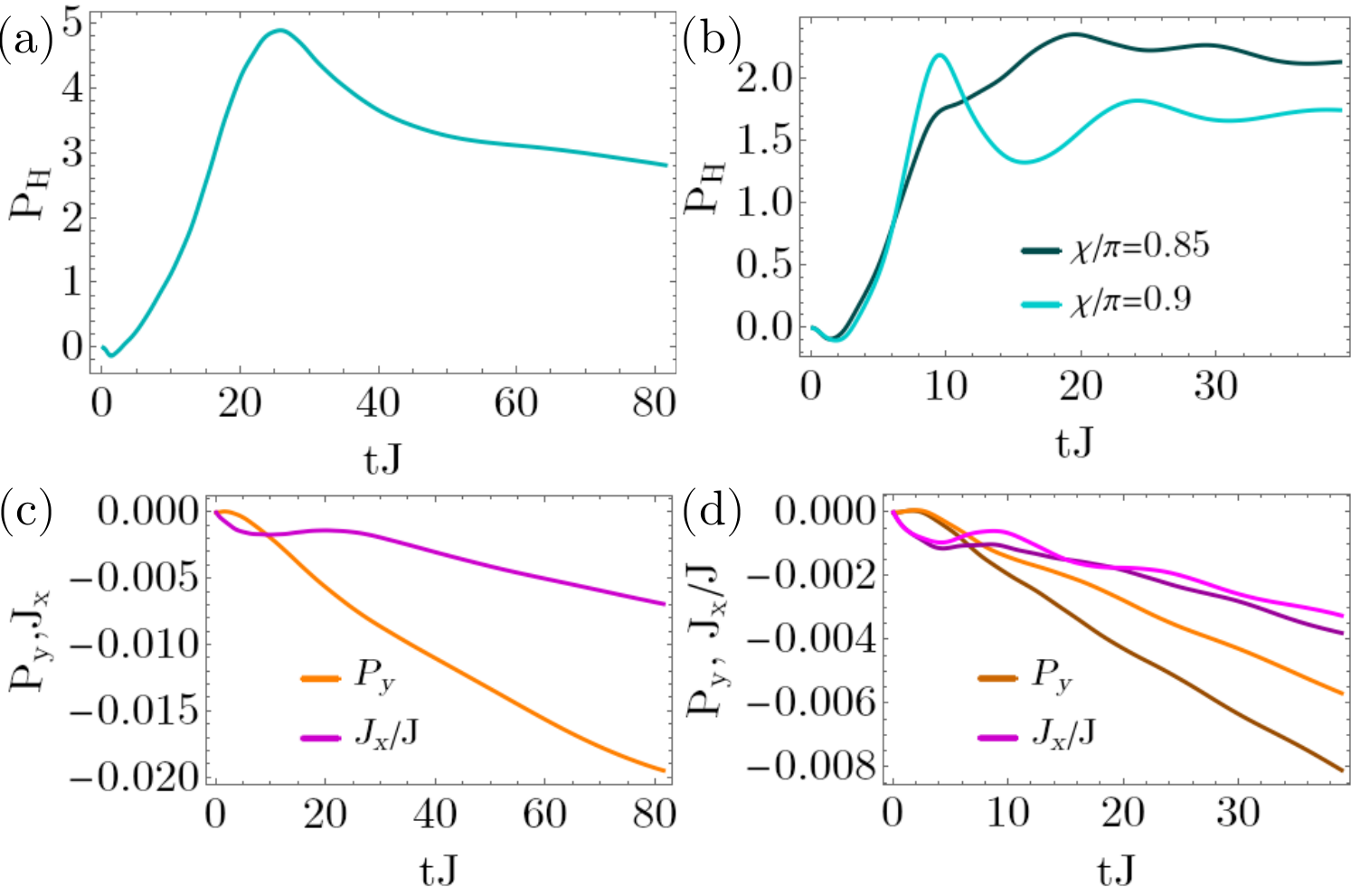}
\caption{\label{fig:pht_vsf2}
(a)-(b) The time evolution of $P_H$ in the vortex superfluid phase for (a) $\chi=0.8\pi$, (b) $\chi=0.85\pi$ and $\chi=0.9$.
(c)-(d) The time evolution of $P_y$ and $\boldsymbol{J}_x$ for (c) $\chi=0.8\pi$, (d) $\chi=0.85\pi$ and $\chi=0.9$. 
In panel (d) the lightest shade of orange and magenta corresponds to the lightest shade of cyan in (b).
The values of $\boldsymbol{J}_x$ and $P_y$ are normalized with the number of rungs.
The parameters used are $J_\|/J=0.8$, $L=90$ rungs, $\rho=0.25$ and $\mu/J=0.001$.
}
\end{figure}

A change of the sign of the value at which $P_H$ stabilizes occurs also in the vortex superfluid, see Fig.~\ref{fig:pht_vsf1} and the main text. 
After the transition from the Meissner phase, occurring at $\chi\approx 0.59\pi$ for $J_\|/J=0.8$, the magnitude of $\langle\langle P_H\rangle\rangle$ decreases and reaches values close to zero around $\chi\gtrsim 0.7\pi$ [Fig.~\ref{fig:pht_vsf1}(a)].
If we continue to increase the flux, $P_y$ changes sign which determines positive values for $P_H$ [Fig.~\ref{fig:pht_vsf1}(b)]. We can observe for $\chi= 0.75\pi$ in Fig.~\ref{fig:pht_vsf1}(a) that initially $P_H$ is negative and only at longer times it becomes positive and that the steady plateau is reached on a longer time scale than for the smaller values of the flux.
The time scales for reaching a plateau are even longer for larger values of the flux, as seen in Fig.~\ref{fig:pht_vsf2}(a) for $\chi=0.8\pi$ and Fig.~\ref{fig:pht_vsf2}(b) for $\chi=0.85\pi,~0.9\pi$.
In particular, for $\chi=0.8\pi$ we obtain one of the largest positive values for $P_H$ throughout the phase diagram, due to the commensurate effects present in the vortex superfluid discussed in the previous sections. For this value of the flux, in Fig.~\ref{fig:pht_vsf2}(a) we observe that $P_H$ increases up to $tJ\approx 25$ and only afterwards decreases towards a saturation value. However, as the saturation only occurs at very long times, the finite size effects play a noticeable role, see Sec.~\ref{sec:ph_linearpot}.

\subsubsection{Hall Polarization in the vortex lattice superfluid phase \label{sec:ph_vlsf}}

\begin{figure}[!hbtp]
\centering
\includegraphics[width=0.49\textwidth]{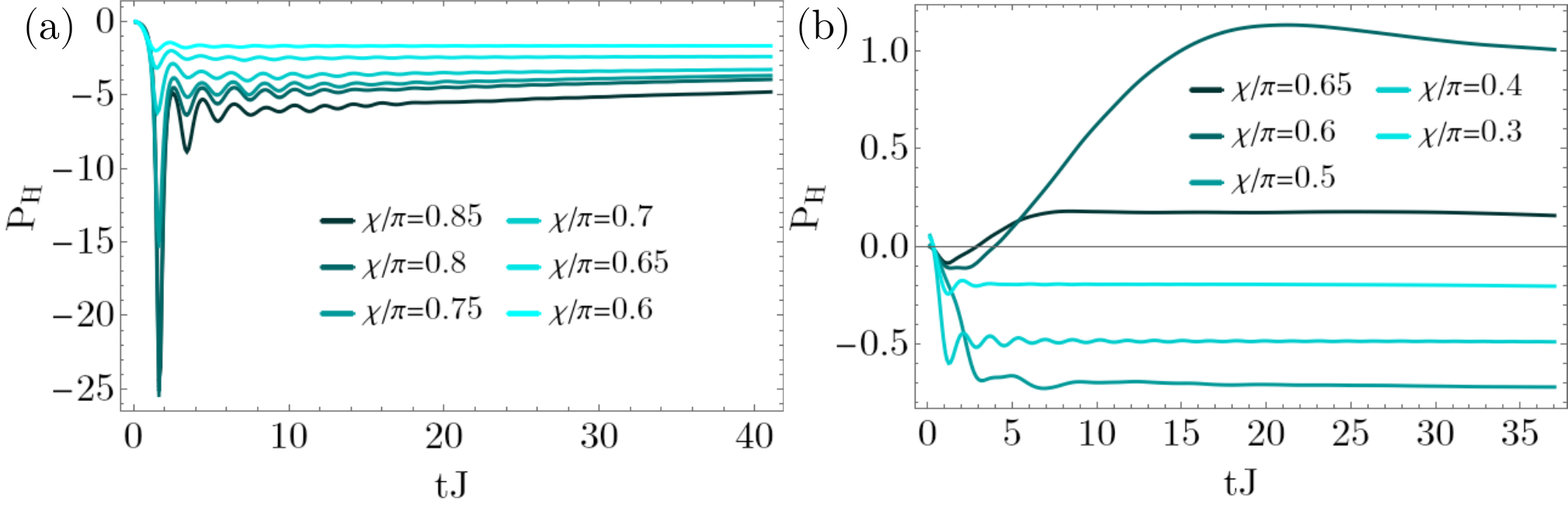}
\caption{\label{fig:pht_vlsf}
(a)-(b) The time evolution of $P_H$ for $\rho=0.4$ and several values of $\chi$ with (a) $J_\|/J=0.6$, and (b) $J_\|/J=1$.
For $J_\|/J=0.6$, a phase transition between the Meissner superfluid and the vortex lattice superfluid with $\rho_v=0.8$ at $\chi\approx0.67\pi$. For $J_\|/J=1$ a succesion of phase transitions from the Meissner superfluid, to the vortex lattice superfluid with $\rho_v=0.6$ and to the vortex superfluid occur at $\chi\approx0.41\pi$ and $\chi\approx0.51\pi$.
The parameters used are $L=90$ rungs and $\mu/J=0.001$.
}
\end{figure}

\begin{figure}[!hbtp]
\centering
\includegraphics[width=0.48\textwidth]{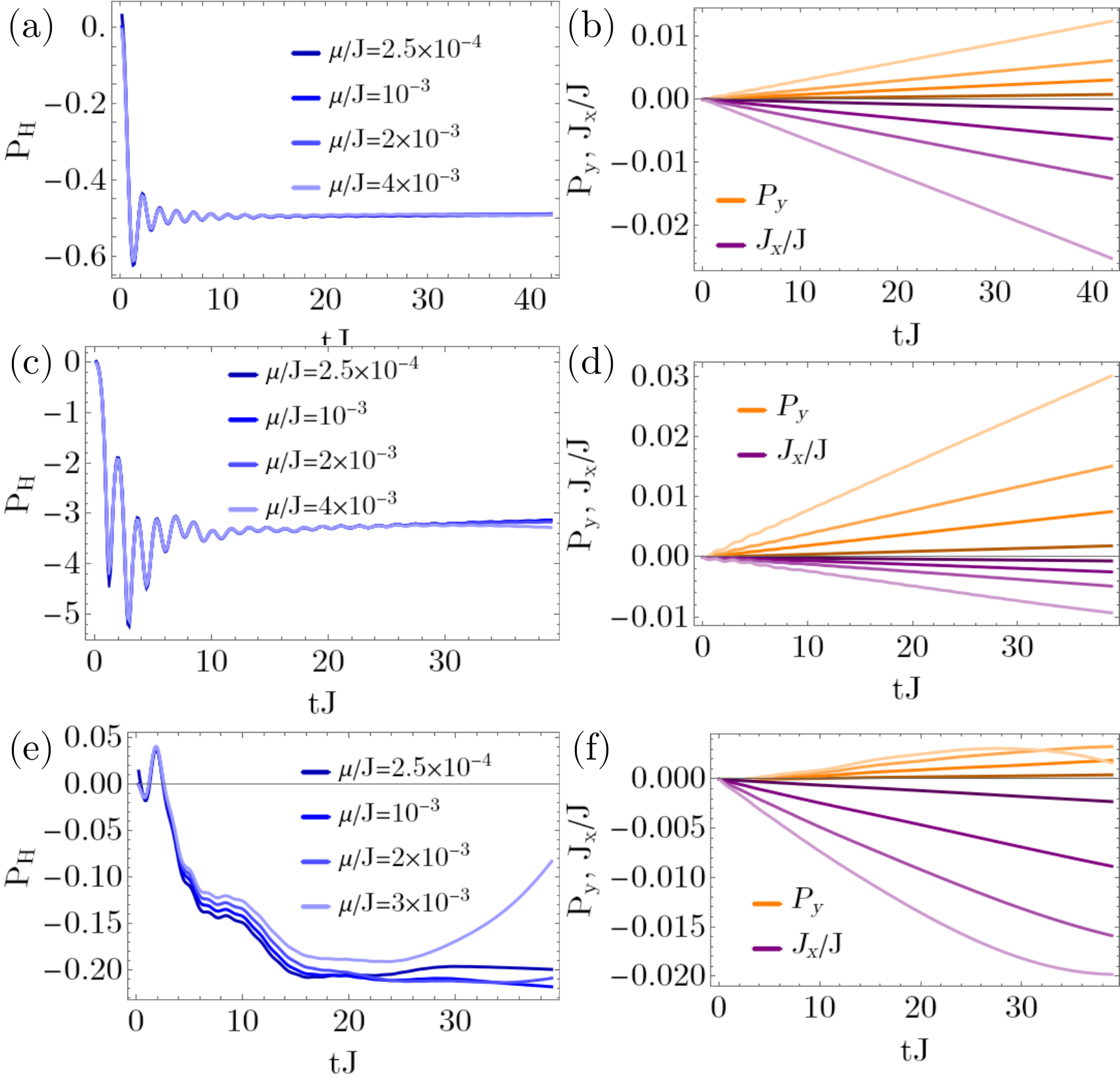}
\caption{\label{fig:mu1}
The time evolution of $P_H$, $P_y$ and $\boldsymbol{J}_x$ for several values of the strength of the linear potential $\mu/J$, for (a)-(b) $J_\|/J=0.2$ and $\chi=0.65\pi$, (c)-(d) $J_\|/J=0.6$ and $\chi=0.6\pi$ in the Meissner superfluid, and (e)-(f) $J_\|/J=0.6$ and $\chi=0.9\pi$ in the biased chiral superfluid.
In the panels on the right depicting $P_y$ and $\boldsymbol{J}_x$, going from the lightest to the darkest shades of orange and purple corresponds to the legends in the panels on the left depicting $P_H$.
The values of $\boldsymbol{J}_x$ and $P_y$ are normalized with the number of rungs.
The parameters used are $L=90$ rungs and $\rho=0.25$.
}
\end{figure}

For an atomic density of $\rho=0.4$ we identified vortex lattice superfluid phase with vortex densities $\rho_v=0.8$ and $\rho_v=0.6$, see Appendix~\ref{appC}, and $\langle\langle P_H\rangle\rangle$ in these phases is shown in Fig.~\ref{fig:PH_density}(b).
In Fig.~\ref{fig:pht_vlsf} we show the time dependence of the Hall polarization in these phases.
We observe in Fig.~\ref{fig:pht_vlsf}(a) for $J_\|/J=0.6$ after we cross the phase boundary at $\chi\approx0.67\pi$ from the Meissner superfluid into the vortex lattice superfluid phase with $\rho_v=0.8$ $P_H(t)$ saturates to large negative values and exhibits a prominent peak at short times, $tJ\approx2$ followed by damped oscillations.
For $J_\|/J=1$ and $\chi=0.5\pi$ in Fig.~\ref{fig:pht_vlsf}(b) we are in the vortex lattice superfluid phase with $\rho_v=0.6$. 
We can see that the dynamics in the vortex lattice superfluid phases does not resemble the one in the vortex superfluid in the region where we have the additional commensurate vortex density related to the density (see Appendix~\ref{appC}), which can be seen for $\chi=0.6\pi$ in Fig.~\ref{fig:pht_vlsf}(b).

\subsubsection{Influence of the linear potential and finite size effects \label{sec:ph_linearpot}}

\begin{figure}[!hbtp]
\centering
\includegraphics[width=0.48\textwidth]{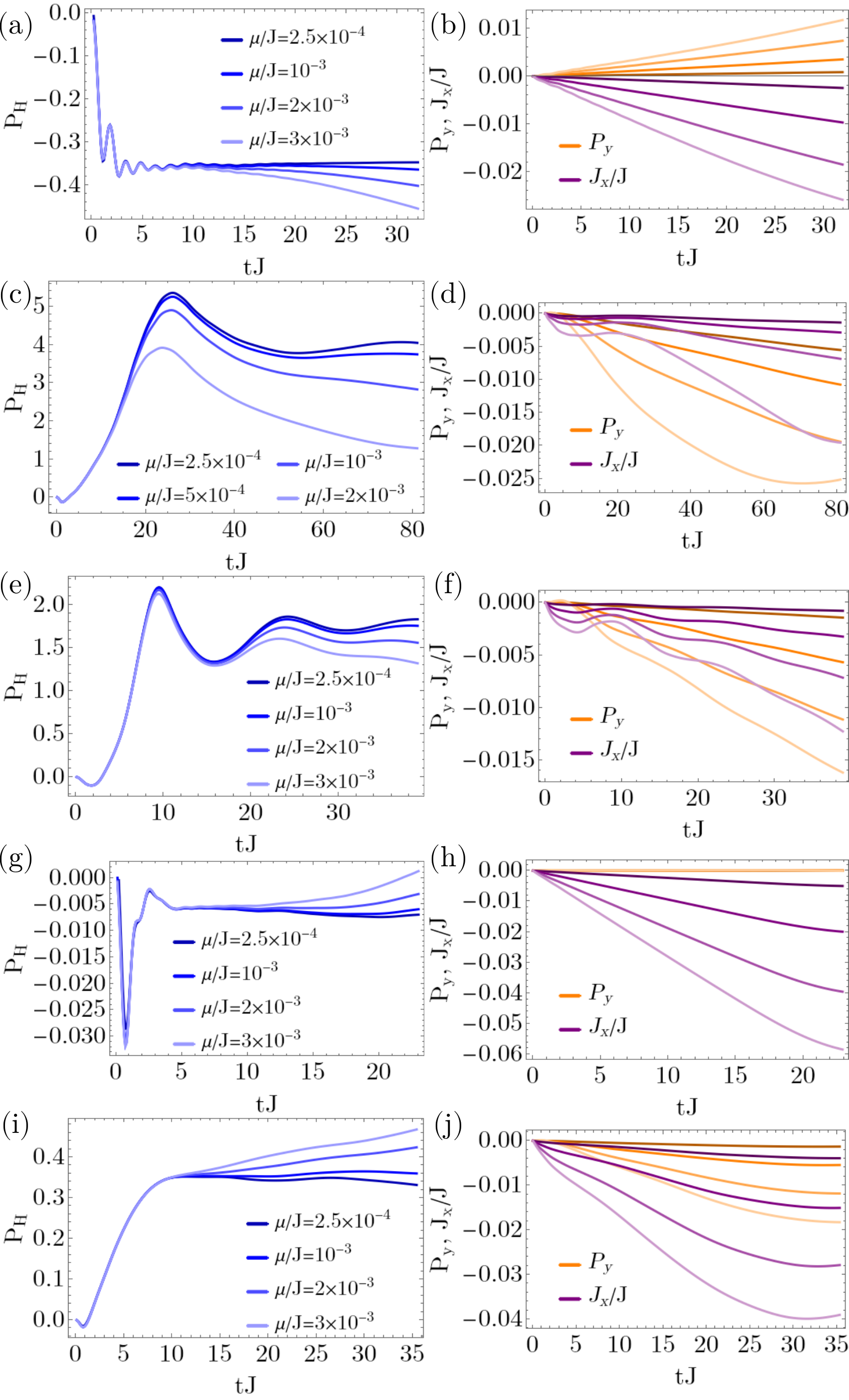}
\caption{\label{fig:mu2}
The time evolution of $P_H$, $P_y$ and $\boldsymbol{J}_x$ for several values of the strength of the linear potential $\mu/J$ in the vortex superfluid phase, for (a)-(b) $J_\|/J=0.8$ and $\chi=0.65\pi$, (c)-(d) $J_\|/J=0.8$ and $\chi=0.8\pi$, (e)-(f) $J_\|/J=0.8$ and $\chi=0.9\pi$, (g)-(h) $J_\|/J=1.3$ and $\chi=0.65\pi$, and (i)-(j) $J_\|/J=1.3$ and $\chi=0.8\pi$.
In the panels on the right depicting $P_y$ and $\boldsymbol{J}_x$, going from the lightest to the darkest shades of orange and purple corresponds to the legends in the panels on the left depicting $P_H$.
The values of $\boldsymbol{J}_x$ and $P_y$ are normalized with the number of rungs.
The parameters used are $L=90$ rungs and $\rho=0.25$.
}
\end{figure}

In this Appendix we analyze the consistency of our results for the dynamics of the Hall polarization under the variation of the strength of the linear potential $\mu$ quenched in the protocol described in the main text.
Furthermore, as the strength of the potential determines the magnitude of the current flowing through the ladder, it can be used also to infer the finite size effects in our results. For stronger potentials, which induce a current $\boldsymbol{J}_x$ growing faster in time, the boundary effects will be noticeable at shorter times, compared to the weaker potentials. 
Thus, we used a value of $\mu/J=10^{-3}$ in the main text which allowed us to capture the long time dynamics of $P_H$.

In the Meissner phase, Fig.~\ref{fig:mu1}(a),(c), we see very little dependence while varying the potential with more than an order of magnitude, from $\mu/J=2.5\times10^{-4}$ to $\mu/J=4\times10^{-3}$.
The results in the biased chiral superfluid are more sensitive to the strength of the potential, for $\mu/J=3\times10^{-3}$ we observe in Fig.~\ref{fig:mu1}(e) that $P_H$ starts increasing from $tJ\approx25$, behavior which is not present for smaller values of the potential and we can attribute to the finite size of the system.
However, for $\mu/J=2.5\times10^{-4}$ to $\mu/J=2\times10^{-3}$ the behavior of $P_H$ is consistent for the different values and only small shifts are present for $tJ\gtrsim8$.

In Fig.~\ref{fig:mu2} in the vortex superfluid phase, we can see that for smaller values of $\mu$ we obtain a steady plateau in $P_H(t)$ for longer times and the point in time where we see deviations from this value is determined by the value of $\mu$.
This holds for different parameters in the vortex phase, as we show in Fig.~\ref{fig:mu2}, which exemplifies the different behaviors we obtained.
In particular, we observe in Fig.~\ref{fig:mu2}(c) and Fig.~\ref{fig:mu2}(e), corresponding to the regime of large positive values of the Hall response, that for lower values of $\mu$ the value of $P_H$ is higher.
In Fig.~\ref{fig:mu2}(c) where we need to go to very long times, $P_H$ for the value we used in the main text of $\mu/J=10^{-3}$ shows deviations which underestimate the saturation value of the Hall polarization, compared to the results for $\mu/J=2.5\times10^{-4}$ and $\mu/J=5\times10^{-4}$. Thus, the features of large positive values of the Hall response discussed in the main text are even more pronounced if one extrapolates the results in the limit of small potential strengths and larger system sizes.

\subsubsection{Identifying phase transitions in the time-dependence of the current \label{sec:current_pt}}

\begin{figure}[!hbtp]
\centering
\includegraphics[width=0.48\textwidth]{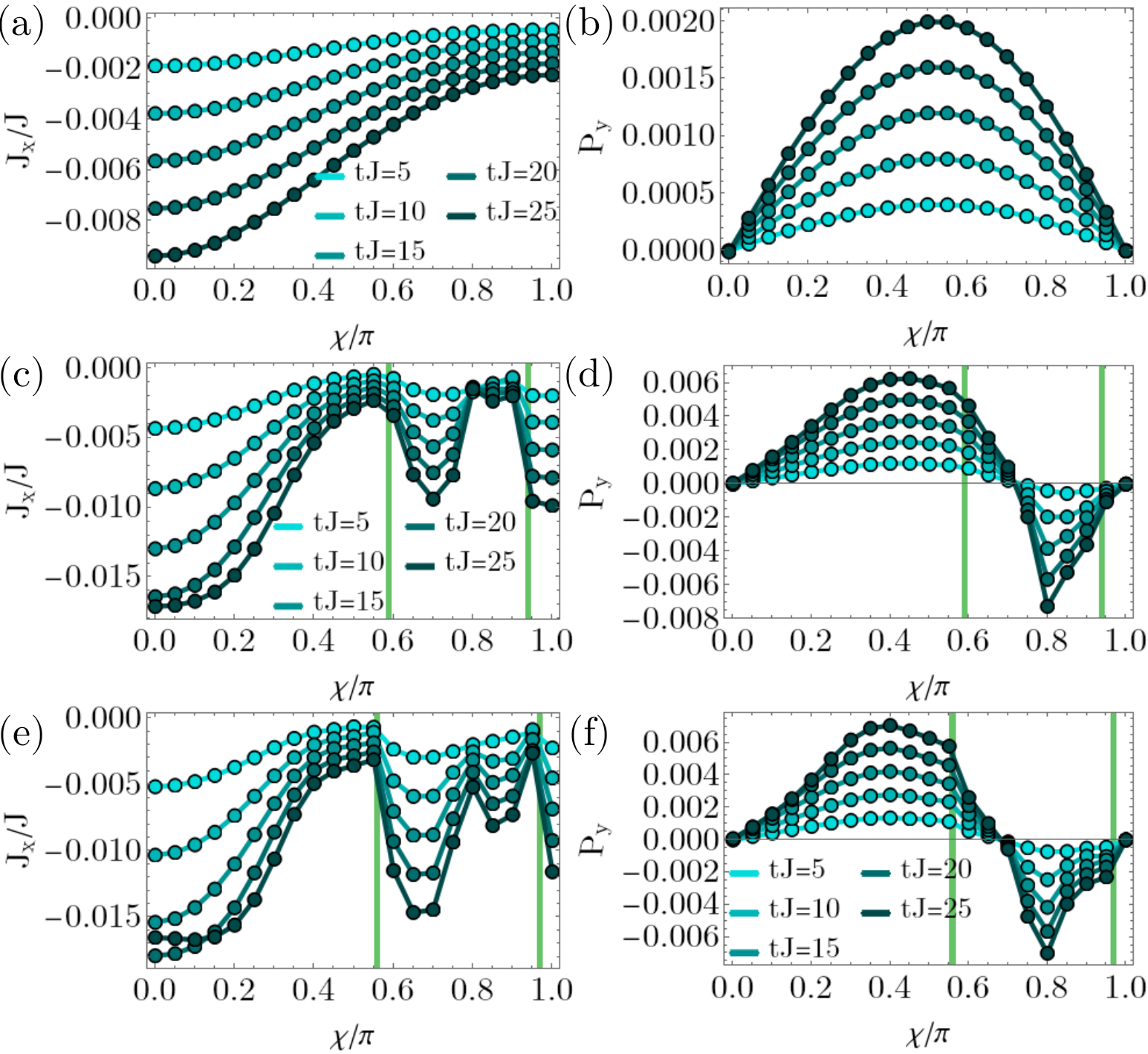}
\caption{\label{fig:current_PT}
The dependence of $P_y$ and $\boldsymbol{J}_x$ as a function of the flux $\chi/\pi$ at different times $tJ$, for (a)-(b) $J_\|/J=0.2$ [corresponding to the parameters in Fig.~\ref{fig:PH_vs_chi}(a)], (c)-(d) $J_\|/J=0.8$ [corresponding to the parameters in Fig.~\ref{fig:PH_vs_chi}(c)], (e)-(f) $J_\|/J=1.0$.
The green vertical lines mark the phase boundaries between M-SF, V-SF and BC-SF phases.
The values of $\boldsymbol{J}_x$ and $P_y$ are normalized with the number of rungs.
The parameters used are $L=90$ rungs, $\rho=0.25$ and $\mu/J=0.001$.}
\end{figure}

In this section, we analyze how the features exhibited by the Hall polarization discussed in Fig.~2 of the main text reflect individually in the behavior of the current $\boldsymbol{J}_x$ and the density imbalance $P_y$.
In Fig.~\ref{fig:current_PT} we plot $\boldsymbol{J}_x$ and $P_y$ as a function of the flux $\chi$ at certain points in time.
For $J_\|/J=0.2$ [Fig.~\ref{fig:current_PT}(a),(b)] the current has a monotonic dependence on the flux and $P_y$ exhibits a single maximum around $\chi\approx0.5\pi$, valid for all shown times. Thus, similar to the behavior of the time-averaged Hall polarization we do not observe any striking feature for $J_\|/J=0.2$ as we are in the Meissner superfluid phase for all values of the flux.
However, if we increase the tunneling amplitude to $J_\|/J=0.8$, where we have a phase transition from M-SF to V-SF, $\chi\approx0.59\pi$, and from V-SF to BC-SF, $\chi\approx0.94\pi$, we observe that the features discussed for $\langle\langle P_H\rangle\rangle$ in the main text are also reflected in the current and density imbalance [Fig.~\ref{fig:current_PT}(c),(d)].
The large negative peak occurring in $\langle\langle P_H\rangle\rangle$ at the Meissner to vortex transition stems for the small magnitude $\boldsymbol{J}_x$ at the phase transition threshold, as seen Fig.~\ref{fig:current_PT}(c). We can further observe that the local maximum of $\boldsymbol{J}_x$ becomes even more apparent at later times, signaling the occurrence of the phase transition.
We obtain similar behavior for $J_\|/J=1.0$ [Fig.~\ref{fig:current_PT}(e),(f)], with $\boldsymbol{J}_x$ showing a small value at the M-SF to V-SF transition, and, interestingly, in the vortex phase. The small values in the vortex phase of the current coincide with the positive peak of the Hall polarization, $\langle\langle P_H\rangle\rangle$, which we associate with the frustration induced commensurability effects (see Sec.~\ref{sec:results}).
While $P_y$ does not seem to be particularly sensitive to the Meissner to vortex transition, the change in sign of the Hall response is due to the sign change of the density imbalance $P_y$ occurring for large values of the flux in Fig.~\ref{fig:current_PT}(f).

\setcounter{equation}{0}
\renewcommand{\theequation}{E.\arabic{equation}}
\setcounter{figure}{0}
\renewcommand{\thefigure}{E\arabic{figure}}

\section{Origin of commensurate vortex density: going between free fermions and hardcore bosons \label{sec:fermions_to_bosons}}

\begin{figure}[!hbtp]
\centering
\includegraphics[width=0.49\textwidth]{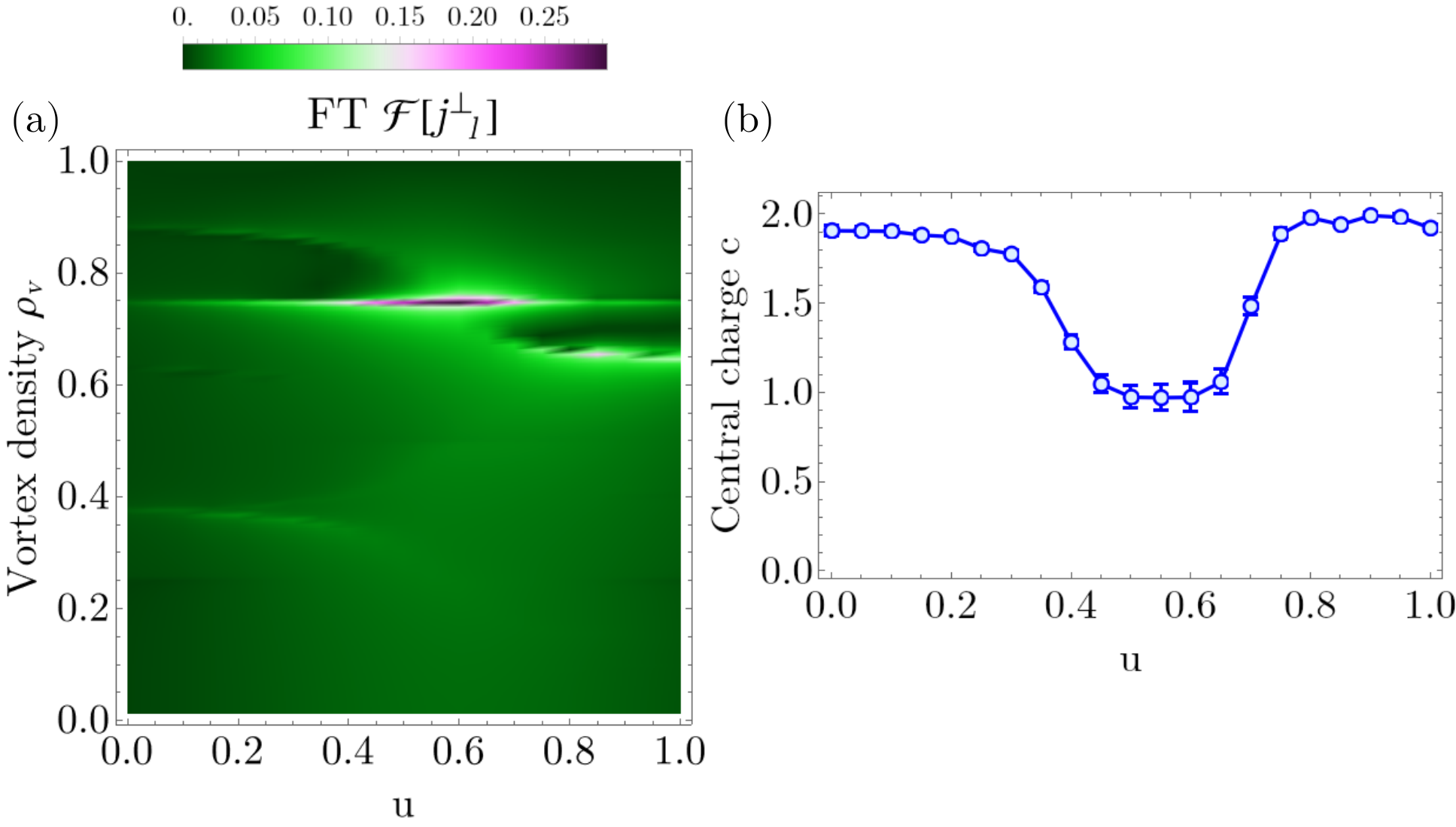}
\caption{\label{fig:fermions_to_bosons}
(a) The Fourier transform of the ground state local rung currents, $j^\perp_j$, and (b) the central charge, $c$, as a function of the interaction $u$, for $J_\|/J=1$, $\chi=0.7\pi$, $\rho=0.25$, $L=120$.
In (a) the vertical axis has been scaled in terms of the vortex density $\rho_v$.
}
\end{figure}

One approach to deal with a model of hardcore bosons is to employ the Jordan-Wigner transformation to spinless fermions 
\begin{align}
\label{eq:JordanWigner}
b_j=\prod_{l=1}^{j-1}e^{i\pi c^\dagger_l c_l}c_j,
\end{align}
where $c_j$ are fermionic operators. In order to use this transformation we first rewrite the Hamiltonian, Eq.~(1) in the main text, on a chain along the rungs of the triangular ladder
\begin{align} 
\label{eq:Hamiltonian_chain}
 H_\text{chain}= &-J \sum_j \left( b^\dagger_{j}b_{j+1} + \text{H.c}. \right)   \\
& -J_\| \sum_{j} \left[ e^{i(-1)^j\chi}b^\dagger_{j}b_{j+2}+ \text{H.c}. \right]. \nonumber
\end{align}
In the language of fermionic operators, using Eq.~(\ref{eq:JordanWigner}), we obtain
\begin{align} 
\label{eq:Hamiltonian_chain_fermions}
 H_\text{f}&=  -J \sum_j \left( c^\dagger_{j}c_{j+1} + \text{H.c}. \right)  \\
&-J_\| \sum_{j} \left[ e^{i(-1)^j\chi}c^\dagger_j(1-2c^\dagger_{j+1}c_{j+1})c_{j+2}+ \text{H.c}. \right] . \nonumber
\end{align}
We observe that due to the hopping term over two sites, stemming from the triangular geometry, the Jordan-Wigner string does not cancel and gives rise to an interaction term. Thus, the hardcore bosons are not mapped to free fermions on the triangular ladder and we obtain an interacting fermionic model.

In order to gauge the importance of the four-fermion interaction term, in the following we vary its strength
\begin{align} 
\label{eq:Hamiltonian_chain_fermions_u}
 H_\text{f}(u)&=  -J \sum_j \left( c^\dagger_{j}c_{j+1} + \text{H.c}. \right)  \\
&-J_\| \sum_{j} \left[ e^{i(-1)^j\chi}c^\dagger_j(1-2uc^\dagger_{j+1}c_{j+1})c_{j+2}+ \text{H.c}. \right]. \nonumber
\end{align}
In the limit of $u=0$ the model reduces to free fermions on a triangular flux ladder, while for $u=1$ we recover the hardcore bosons.

In Fig.~\ref{fig:fermions_to_bosons} we study the nature of the ground state as a function of $u$. At intermediate values of the interaction, $0.4\lesssim u\lesssim0.7$, we observe a strong peak in the Fourier transform of the local rung currents at a vortex density of $\rho_v=0.75$.
This fact together with the value of the central charge of $c\approx 1$ in this regime points towards the existence of a vortex lattice phase in the interacting fermion model.
Interestingly, the peak at $\rho_v=0.75$ does not vanish when we enter the gapless $c=2$ phases, either towards free fermions, or hardcore bosons.
Thus, this suggests that the origin of the vortex density commensurate with the filling stems from the interplay of interactions and the triangular geometry, due to the proximity of the vortex lattice with $\rho_v=0.75$.

We note that the local rung currents have an analogous definition for fermions and hardcore bosons, as they connect neighboring sites in the chain representation the Jordan-Wigner string vanishes. This makes them a relevant observable to analyze as we vary $u$ to go from free fermions to hardcore bosons in our system.

\end{document}